\documentclass[
]{jss}

\usepackage[utf8]{inputenc}

\author{
Etienne Côme\\COSYS - GRETTIA\\
Université Gustave Eiffel\\
\strut \\
F-77454 Marne-la-Vallée \And Nicolas Jouvin\\UMR 518 MIA Paris-Saclay\\
INRAE, AgroParisTech\\
Université Paris-Saclay\\
16 rue claude bernard, 75005 Paris
}
\title{\pkg{greed}: An \proglang{R} Package for Model-Based Clustering
by Greedy Maximization of the Integrated Classification Likelihood}

\Plainauthor{Etienne Côme, Nicolas Jouvin}
\Plaintitle{greed: An R Package for Model-Based Clustering by Greedy
Maximization of the Integrated Classification Likelihood}
\Shorttitle{\pkg{greed}: Model-Based Clustering with the Integrated
Classification Likelihood}

\Abstract{
The \pkg{greed} package implements the general and flexible framework of
\citet{come2021hierarchical} for model-based clustering in the
\proglang{R} language. Based on the direct maximization of the exact
Integrated Classification Likelihood with respect to the partition, it
allows jointly performing clustering and selection of the number of
groups. This combinatorial problem is handled through an efficient
hybrid genetic algorithm, while a final hierarchical step allows
accessing coarser partitions and extract an ordering of the clusters.
This methodology is applicable in a wide variety of latent variable
models and, hence, can handle various data types as well as
heterogeneous data. Classical models for continuous, count, categorical
and graph data are implemented, and new models may be incorporated
thanks to S4 class abstraction. This paper introduces the package, the
design choices that guided its development and illustrates its usage on
practical use-cases.
}

\Plainkeywords{R, model-based, clustering, hierarchical, classification
likelihood, genetic algorithms}

\Address{
    Etienne Côme\\
    COSYS - GRETTIA\\
Université Gustave Eiffel\\
\strut \\
F-77454 Marne-la-Vallée\\
    Marne La Vallée\\
  E-mail: \email{etienne.come@univ-eiffel.fr}\\
  
      Nicolas Jouvin\\
    UMR 518 MIA Paris-Saclay\\
INRAE, AgroParisTech\\
Université Paris-Saclay\\
16 rue claude bernard, 75005 Paris\\
    Paris\\
  E-mail: \email{nicolas.jouvin@inrae.fr}\\
  
  }

\providecommand{\tightlist}{%
  \setlength{\itemsep}{0pt}\setlength{\parskip}{0pt}}

\usepackage{longtable}
\usepackage{array}
\usepackage{amsmath}
\usepackage{bm}
\usepackage{amsfonts} 
\usepackage{amssymb} 
\usepackage[dvipsnames]{xcolor}
\usepackage{hyperref}
\usepackage{cleveref}

\usepackage{array}

\newcolumntype{L}[1]{>{\raggedright\let\newline\\\arraybackslash\hspace{0pt}}m{#1}}
\newcolumntype{C}[1]{>{\centering\let\newline\\\arraybackslash\hspace{0pt}}m{#1}}
\newcolumntype{R}[1]{>{\raggedleft\let\newline\\\arraybackslash\hspace{0pt}}m{#1}}

\newcommand*\diff{\mathop{}\!\mathrm{d}} 

\DeclareMathOperator*{\argmax}{arg\,max}

\DeclareMathOperator{\ICLex}{ICL_{\textit{ex}}} 
\DeclareMathOperator{\ICL}{ICL}

\DeclareMathOperator{\Dir}{Dir}

\newcommand{\Mult}{\mathcal{M}}
\newcommand{\Gaussian}{\mathcal{N}}

\newcommand{\Partition}{\mathcal{P}}

\bmdefine{\bC}{C}
\newcommand{\p}{p}

\newcommand{\nb}{n}
\newcommand{\K}{K}

\bmdefine{\Obs}{\bm{X}}
\bmdefine{\obs}{\bm{x}}
\newcommand{\rawobs}{x}
\bmdefine{\Clust}{\bm{Z}}
\bmdefine{\clust}{\bm{z}}
\newcommand{\rawclust}{z}
\bmdefine{\scores}{\obs}

\bmdefine{\param}{\bm{\theta}}

\bmdefine{\globalparam}{\vartheta}

\bmdefine{\bPi}{\bm{\pi}}

\bmdefine{\bmean}{\bm{m}}
\bmdefine{\bS}{\bm{S}}
\bmdefine{\bD}{\bm{D}}
\bmdefine{\bDelta}{\Delta}

\bmdefine{\bBeta}{\beta}

\bmdefine{\balpha}{\bm{\alpha}}

\bmdefine{\Clustrow}{\bm{Z}_{r}}
\bmdefine{\clustrow}{\bm{z}}

\bmdefine{\Clustcol}{\bm{Z}_{c}}
\bmdefine{\clustcol}{\bm{z}}

\bmdefine{\bPirow}{\bm{\pi}_{r}}
\bmdefine{\bPicol}{\bm{\pi}_{c}}

\renewcommand{\Clustrow}{\bm{Z}^{r}}
\renewcommand{\clustrow}{\bm{z}^{r}}

\renewcommand{\Clustcol}{\bm{Z}^{c}}
\renewcommand{\clustcol}{\bm{z}^{c}}

\renewcommand{\bPirow}{\bm{\pi}^{r}}
\renewcommand{\bPicol}{\bm{\pi}^{c}}

\begin{document}

\hypertarget{introduction}{%
\section{Introduction}\label{introduction}}

Clustering consists in the unsupervised task of grouping a set of
objects into distinct groups or \textit{clusters}. Unveiling relevant
structure in datasets, it holds an important part in modern data
analysis, with a wide range of applications involving data of different
nature. Grounded on a statistical approach, model-based clustering
provides a flexible method capable of handling the variety and
complexity of modern data such as continuous, count, graphs or mixed
data in a common framework using finite mixtures or stochastic block
models \citep{bouveyron2019model}. The \pkg{greed} \proglang{R} package,
introduced in this paper, builds on this general framework and provides
a generic method for clustering various types of data based on the
maximization of the Integrated Classification Likelihood
\citep{come2021hierarchical}. The method is versatile and thanks to the
exploration capabilities of an hybrid genetic algorithm, it does not
rely on a carefully chosen - and potentially costly and model dependent
- initialization procedures. Throughout the paper, we will discuss
specific instances of models and give suitable references for related
works and packages for each of the considered data type: count,
continuous, mixed-type and graph data.

The clustering problem has been the focus of a lot of attention in the
last decades, and the \proglang{R} \citep{Rcore} community have been a
driving
force\footnote{A snapshot highlighting the numerous contributions made is available on a dedicated CRAN ``Task Views'' page on clustering: \url{https://cran.r-project.org/web/views/Cluster.html}}
with the development of many packages taking advantage of the
\proglang{R} programming language capacities in handling and visualizing
data. These packages can be divided into three main categories:

\begin{enumerate}
 \item On the one hand, there are packages implementing distance-based clustering, relying on some \textit{ad hoc} notion of similarity between observations. Working with vector-valued observations, the well-known k-means \citep{macqueen1967some} algorithm is implemented in the \pkg{stats} and \pkg{ClusterR} \citep{ClusterR} packages for Euclidean distances. Refined similarity metrics for k-means such as kernels are available in the \pkg{kernlab} package \citep{kernlab}, together with an implementation of the spectral clustering algorithm \citep{ng2002spectral}. Other partitional methods working with k-medoids and arbitrary distance functions are implemented in the \pkg{cluster} package \citep{clusterpackage}. Finally, the \pkg{clustMixType} \citep{clustMixType} package allows to cluster mixed type data composed of numeric and categorical variables. On the side of graph clustering algorithms, the well-known Louvain algorithm \citep{blondel2008fast} based on modularity maximization is implemented in the \pkg{igraph} \citep{igraph}, and the spectral clustering algorithm is also very popular for clustering binary or weighted adjacency matrices. 
 \item On the other hand, packages implementing model-based clustering fit a probabilistic model to the data. We distinguish between two main strategies for statistical inference:
 \begin{enumerate}
 \item \textit{Frequentist}: The first strategy casts clustering as a parameter estimation problem, traditionally dealt with via maximum likelihood, and most of the time solved with the help of some (possibly variational) EM procedure. The clustering comes as a byproduct since a partition can then be obtained via its posterior distribution given the observation and the point estimates. This approach is implemented in various \proglang{R} packages such as \pkg{Mclust} \citep{Mclust} for Gaussian mixture models, \pkg{mixtools} \citep{mixtools} for multinomial mixtures, \pkg{poLCA} \citep{poLCA} for mixture of categorical distributions, and \pkg{blockmodels} \citep{blockmodels} and \pkg{sbm} \citep{sbm} for the simple and bipartite stochastic block models (SBM). The \pkg{Rmixmod} \citep{Rmixmod} package also provides an \proglang{R} interface with the MIXMOD software to fit Gaussian or categorical mixtures for continuous or count data clustering. Somewhat closer to our contribution, the \pkg{flexmix} package \citep{flexmix} also proposes a general framework for fitting mixture models on, possibly heterogeneous, multivariate data using maximum likelihood and an EM procedure. However, it does not cover graph data clustering with SBMs since the latter cannot be cast as simple mixture models and cannot be estimated through regular EM procedures. Their approach is extensible, and new mixture models may be implemented via an S4 class representation of their M-step. While the algorithms strongly differ, our \pkg{greed} package proposes a similar interface, with the possibility to implement new models as S4 classes without impacting the main functions and API.
 
 \item \textit{Bayesian}: The second strategy relies on Bayesian inference, seeking to accurately estimate the posterior distribution of the model parameters and cluster memberships, usually via Markov chain Monte Carlo methods \citep[MCMC,][]{robert2013monte}. Some of these methods allow inferring the number of clusters through a nonparametric Dirichlet process prior on the group proportions. The main packages implementing Bayesian model-based clustering are \pkg{rjags} \citep{rjags} and \pkg{rstan} \citep{RStan}, respectively \proglang{R} interface to the JAGS and STAN softwares, each having dedicated modules for finite mixture modelling. Other packages directly focus on mixture modelling such as \pkg{Bmix} \citep{Bmix}, \pkg{bmixture} \citep{bmixture} or \pkg{IMIFA} \citep{IMIFA}. Note that, in addition to heavy computations, the two principal difficulties of MCMC inference for clustering are label switching and the resulting multimodality of the posterior. 
 \end{enumerate}
\end{enumerate}

Beyond the \proglang{R} community, most of the aforementioned methods
are implemented in popular \proglang{Python} modules. Without being
exhaustive, we mention the \pkg{scikit-learn} module
\citep{scikit-learn} implementing standard algorithms for multivariate
data clustering in its \textit{cluster} submodule, and the
\pkg{graph-tool} module \citep{graphtool} which implements Bayesian
inference for the stochastic block model and its variants.

Our approach lies in between the frequentist and Bayesian approaches,
which, while efficient, can be computationally intensive as they require
to explore a high-dimensional parameter space, and possibly perform
model selection by running the inference procedure on a grid of values
for \(K\) \textit{i.e.} the number of clusters. A contrario, we focus on
clustering rather than points or posterior estimates of the model
parameters, by maximizing an \textit{exact} version of the Integrated
Classification Likelihood \citep[ICL,][]{Biernacki2000} directly with
respect to the partition. Originally used as a model selection criterion
in frequentist settings, the exact ICL is rooted in the Bayesian
framework and consists in the joint distribution of the observations and
the partition with all other parameters analytically integrated out by a
suitable choice of prior distributions. The genetic algorithm of
\citet{come2021hierarchical} then solves the highly combinatorial
problem of maximizing this objective function with respect to the
partition, allowing \(K\) to be variable, hence performing clustering
and model selection altogether while avoiding grid procedures.

In a complementary step, the \pkg{greed} package implements a
model-based hierarchical method based on the maximization of a modified
ICL criterion. Starting from an initial solution provided by the genetic
algorithm, it extracts a hierarchy of nested partitions in an ascendant
fashion, by finding the best merge according to this new criterion,
similarly to the \code{hclust()} function of the \pkg{stats} package.
This second step unveils hierarchical structures present in the data,
which is useful both for interpretation and visualization
\citep[Section 4]{everitt2011cluster}, while also giving a dendrogram
representation of the hierarchy along with a meaningful pseudo-ordering
of the clusters.

The paper is organized in the following way. \Cref{sec:theory} briefly
introduces the unified statistical framework considered by
\citet{come2021hierarchical} and gives specific instances of models,
namely finite mixtures and stochastic block models. Then,
\Cref{sec:algo} describes the genetic and hierarchical algorithms along
with practical package implementation details. Finally,
\Cref{sec:guidelines} presents an overview of the package's API and
practical guidelines, while \Cref{sec:applications} demonstrates its
flexibility on practical use-cases, with real-data studies for
continuous, categorical, graph and mixed-type data.

\hypertarget{model-based-clustering}{%
\section{Model-based clustering}\label{model-based-clustering}}

\label{sec:theory} We begin with a brief reminder on model-based
approaches to clustering. We will denote as \(\Obs\) the set of
observations, which can be a collection of \(n\) vectors in
\(\mathbb{R}^p\) in the case of mixture modelling, or an \(n \times n\)
adjacency matrix in the case of graph clustering. The unobserved
partition of \(\{1, \ldots, n\}\) will be denoted as
\(\Clust = \{ \clust_1, \ldots, \clust_n \}\) where \(\clust_i\) is a
binary vector of size \(K\) indicating cluster membership of object
\(i\).

\hypertarget{discrete-latent-variables-models}{%
\subsection{Discrete latent variables
models}\label{discrete-latent-variables-models}}

Model-based clustering can be decomposed in a two-stage generative
process. First, the partition \(\Clust\) is drawn from a product of
multinomial distributions with parameter \(\bPi\), the latter
quantifying the a priori probability to belong to each of the \(K\)
groups. Second, observations are drawn conditionally on the partition,
according to some parametric distribution with parameters \(\param\)
depending on the clusters assignments.

Depending on the context, observations may be continuous or discrete
vectors in dimension \(p\), or edges in a graph. Thus, there are a
variety of \textit{observational models} that can be handled which such
an approach. All the models handled by the \pkg{greed} package share the
common hypothesis of \textit{conditional independence} given the full
partition \(\Clust\): \begin{align}
    \label{HCICL:eq:DLVM}
    \p(\Obs, \Clust\mid \bPi, \param) = 
    \prod_{\clust \in \Clust} \p(\clust \mid \bPi) \underset{\text{conditional independence}}{\underbrace{\prod_{\obs \in \Obs}\p(\obs \mid \Clust, \param)}},
    \end{align} Such models are called discrete latent variables models
\citep[DLVMs,][]{come2021hierarchical} and popular instances for
multivariate data and graph clustering are given in
\Cref{tab:ObservationalModels}. For the sake of brevity, we do not
detail co-clustering with latent block models (LBM) although it fits the
definition of DLVMs and is implemented in \pkg{greed}.

As discussed above, standard frequentist approaches casts clustering as
a parameter estimation problem, traditionally dealt with a
maximum-likelihood approach solved with the help of some (variational)
EM procedure in most cases. The clustering comes as a byproduct since a
partition can then be obtained via the posterior distribution of
\(\Clust \mid \Obs, \hat{\bPi}, \hat{\param}\). This approach is
implemented in various \proglang{R} packages such as \pkg{Mclust}
\citep{Mclust} for Gaussian mixture models, \pkg{mixtools}
\citep{mixtools} for multinomial mixtures, \pkg{flexmix} \citep{flexmix}
for general finite mixture models or \pkg{blockmodels}
\citep{blockmodels} for the stochastic block model.

\begin{table}[!ht]
  \renewcommand{\arraystretch}{1.5}
    \centering
    \begin{tabular}{|l|l|l|}
        \hline
        Model name & Observation type & Observational model: $\Obs \mid \Clust$ \\
        \hline
        \hline
        GMM & Multivariate continuous: $\Obs \in \mathbb{R}^{n\times p}$ & $\obs_{i} \mid \rawclust_{ik}=1, \param_{k} \sim \Gaussian_p(\obs_{i} \mid \bmean_{k}, \bS_k)$ \\
        \hline
        MoM & Multivariate discrete: $\Obs \in \mathbb{N}^{n\times p}$ & $\obs_{i} \mid \rawclust_{ik}=1, \param_{k} \sim \Mult_p(\obs_{i} \mid \param_k)$ \\
        \hline
        SBM &  Graphs: $\Obs \in \mathbb{R}^{n\times n}$  &  $\rawobs_{ij} \mid \rawclust_{ik}\rawclust_{jl}=1, \param_{kl} \sim \p(\rawobs_{ij} \mid \param_{kl})$    \\
        \hline
    \end{tabular}
  \caption{\label{tab:ObservationalModels} Some standard observational models handled by the \pkg{greed} package: Gaussian mixture model (GMM),  mixture of multinomial (MoM) and stochastic block models (SBM) for (possibly weighted) graphs. See \Cref{tab:obsmodS4} for a complete list of implemented DLVMs.}

\end{table}

\hypertarget{exact-integrated-classification-likelihood}{%
\subsection{Exact Integrated Classification
Likelihood}\label{exact-integrated-classification-likelihood}}

Having derived an estimate of the model parameters, the choice of the
number of clusters \(K\) is usually delayed post-inference as a
\textit{model selection} problem
\citep[Chapter 7]{fruhwirth2019handbook}. There exists a large variety
of model selection criteria, generally involving a penalized likelihood,
such as the Akaike Information Criterion \citep[AIC,][]{akaike1974new}
or the Bayesian Information Criterion
\citep[BIC,][]{schwarz1978estimating} which are commonly implemented in
routine packages. Note that, in this framework, performing clustering
and model selection requires to estimate the parameters for a grid of
models indexed by \(K\), which can be quite cumbersome, even for
moderate size problems, as the computational complexity typically grows
with \(K\).

Hereafter, we focus on the ICL of \citet{Biernacki2000}, a widely used
model selection criterion in the clustering context. Adopting a Bayesian
point of view on the model parameters \(\param\) and \(\bPi\), it
consists in integrating the latter out hence naturally penalizing for
complex models. Introducing a factorized prior
\(\p(\param, \bPi) = \p(\param \mid \bBeta) \p(\bPi \mid \balpha)\) with
respective hyperparameters \(\bBeta\) and \(\balpha\), the \(\ICL\)
writes as: \begin{equation}
\label{eq:ICL}
\ICL(\Clust; \balpha, \bBeta)  =  \log \int \p (\Obs \mid \param, \Clust) \p(\param \mid \bBeta) \diff \param + \log \int \p (\Clust \mid \bPi ) \p(\bPi \mid \balpha) \diff \bPi.
\end{equation} Such integrals are typically non-analytic and usually
approximated via Laplace method, leading to a penalized likelihood
criterion \textit{à la} BIC.

However, it is possible to derive exact version of the ICL by choosing
conjugate priors for the model parameters \(\p(\param \mid \bBeta)\) and
\(\p(\bPi \mid \balpha)\): \begin{equation}
\label{eq:ICLex}
\begin{aligned}
\ICLex(\Clust; \balpha, \bBeta) 
& =  \log \p (\Obs \mid \Clust, \bBeta) + \log \p (\Clust \mid \balpha).
\end{aligned}
\end{equation} The first term of the right-hand side is model-dependent,
and must be computed on a case-by-case
basis\footnote{Detailed computations of $\ICLex$ for the considered models can be found in the Supplementary Materials of \citet{come2021hierarchical}}
as it depends on the prior on the observational model parameters
\(\param\). However, putting a symmetric Dirichlet prior
\({\p(\bPi) = \Dir(\bPi \mid \bm{\alpha} = (\alpha, \ldots, \alpha))}\)
over the group proportions, the second term is universal to any DLVM and
is derived thanks to Dirichlet-Multinomial conjugacy: \begin{equation}
\label{eq:ICLexdec}
\ICLex(\Clust; \alpha,\bBeta) = \log \p(\Obs \mid \Clust, \bBeta)  +  \log\left( \dfrac{\Gamma(\K \alpha)\prod\limits_{k=1}^{\K} \Gamma(\alpha + n_k)}{\Gamma(\alpha)^\K  \Gamma(\nb + \alpha\, \K)}\right) ,
\end{equation} where \(n_k\) is the number of individuals in cluster
\(k\).

\hypertarget{algorithms}{%
\section{Algorithms}\label{algorithms}}

\label{sec:algo}

\hypertarget{greedy-maximization-of-the-exact-icl}{%
\subsection{Greedy maximization of the exact
ICL}\label{greedy-maximization-of-the-exact-icl}}

In contrast to inference-based approaches, the \pkg{greed} package
bypasses the statistical inference step and focuses on the clustering
objective of jointly finding some \textit{hard}-partitioning \(\Clust\)
as well as its number of clusters \(K\). This is done by solving the
maximization of \(\ICLex\) with respect to both \((\Clust, K)\):
\begin{equation}
\label{eq:ICLoptim}
    \Clust^{(K^\star)} \in \argmax_{K, \Clust} \ICLex(\Clust, K),
\end{equation} the \pkg{greed} package implements several algorithms
tackling this problem. Naturally, due to the highly multimodal and
combinatorial nature of this discrete problem, the proposed algorithms
are not guaranteed to converge to a global maximum, but rather to
efficiently explore the space of solutions to find relevant local optima
at a reasonable computational time and cost.

The default algorithm used by the \pkg{greed} package is a hybrid
genetic algorithm (GA), introduced in detail in
\citep{come2021hierarchical}. Standard GAs evolve a population of
solutions by selecting some of the most promising ones, crossing them,
and possibly mutating them until a specified number of generations or
some stopping criterion is reached. Their capacity to efficiently
explore huge space of solutions makes them ideal candidate for discrete
combinatorial optimization problems such as in \Cref{eq:ICLoptim}.
Moreover, in order to improve the exploitation capacity of GAs around
local optima, one may hybrid them with efficient local search algorithms
\citep[see][Chap. 10]{Eiben2003} such as greedy heuristics based on
swaps and merge moves. This is the default approach implemented in the
\pkg{greed} package, and we detail its most central features as a GA:
solution representation and the recombination, mutation and selection
operators.

\paragraph{Solutions as partitions.}

Since the \(\ICLex\) function is invariant under permutations of the
cluster indexes, the integer encoding representation in \(\Clust\) is
redundant for our optimization problem. This fact is also known as the
label switching problem and have an important impact on the design of
the recombination (crossover) operator for GA. Indeed, simple
recombination operators based on crossover points will not consider this
particularity and will completely break the structure of the solution
\citep{Hruschka2009}, leading to slow evolution of the population of
solutions. One solution to circumvent this issue is to define the
crossover and mutation operators directly over the space of partitions
\(\{1,\hdots, \nb\}\) into \(\K\) clusters with a variable \(\K\),
denoted as \(\Partition = \{\bC_1,...,\bC_{\K}\}\). Such operators will
not suffer from the label switching problems and will therefore not
break the structures already found. The hybrid GA used in \pkg{greed} is
based on such operators.

\paragraph{Combining two partitions: the crossover operator.}

The crossover operator, recombining two ``parents'' solutions into a new
``child'' one, is based on the \textit{cross partition} operator. The
cross partition of two partitions is simply the partition built by
considering all the possible intersections between the elements of the
two partitions being crossed. More formally:

\begin{equation}
\label{eq:CrossPartition}
\mathcal{P}^1\times\mathcal{P}^2 := \{\mathbf{C}^1_i \cap \mathbf{C}^2_j \,,\, \forall i\in \{1,...,|\mathcal{P}^1|\}, j\in \{1,...,|\mathcal{P}^2|\}\}\setminus \emptyset,
\end{equation}

with \(\mathcal{P}^1=\{\mathbf{C}_1^1,...,\mathbf{C}^1_{K_1}\}\) and
\(\mathcal{P}^2=\{\mathbf{C}_1^2,...,\mathbf{C}^2_{K_2}\}\) two
partitions of \(\{1,..., n\}\). This operator produces a new solution
which is a refinement of both solutions being crossed, with at most
\(|\mathcal{P}^1|.|\mathcal{P}^2|\) clusters. In practice, the two
solutions being crossed will agree on some clusters and the number of
new clusters after crossover will be smaller. It also defines the
coarser clustering which can lead to either \(\mathcal{P}^1\) or
\(\mathcal{P}^2\) using merge operations, \textit{i.e.} their first
common ancestor in the partition lattice. Thus, in case of under-fitting
of both parent partitions, crossing alone may already improve the
solution. Finally, while this crossing may create superfluous clusters
when done near local maxima of \(\ICLex\), the greedy local search based
on merges used in the hybrid GA efficiently removes these extra
clusters.

\paragraph{Mutation \& selection.}

The remaining aspects to set up concern the selection procedures and the
mutation operators. For the selection process, the hybrid GA algorithm
uses a classical rank-based selection policy
\citep[see][p. 81-82]{Eiben2003}. In this scheme, at each step, the
\(V\) solutions selected for building the next generation are selected
according to a probability proportional to their rank in terms of
\(\ICLex\). Eventually, regarding the mutation operator, the hybrid GA
algorithm splits a random cluster in two at random. Indeed, while the
greedy heuristic, consisting in swaps and merges, can decrease the
complexity of the solutions, it is unable to refine a partition. Such a
mutation, along with the recombination operator, will help the
exploration of candidate solutions with more clusters.

Eventually, note that the first generation of candidates build by the
hybrid-GA algorithm are constructed using a simple greedy swap algorithm
from totally random starting partitions. Greedy hill-climbing heuristics
are therefore used at initialization (swaps) and after each
recombination (merges) and mutations (swaps).

Finally, the \pkg{greed} package also comes along with three other
optimization algorithms that the user may decide to use for comparison
or if suitable:

\begin{itemize}
   \item A classical genetic algorithm without hybridization with greedy local search.
 \item A classical greedy hill climbing (swap and merge) algorithm with
    multiple random starting partitions \citep[see \textit{e.g.}][for the SBM]{come2015} .
 \item A classical greedy hill climbing algorithm \citep{come2015} with one \textit{seeded} starting point, the \textit{seed} being produced by a model-dependent heuristic (over-segmented $K$-means for GMM as an example).
\end{itemize}

\hypertarget{hierarchical-clustering-and-cluster-ordering}{%
\subsection{Hierarchical clustering and cluster
ordering}\label{hierarchical-clustering-and-cluster-ordering}}

Once a solution with a number \(K^*\) of clusters have been found by one
of the previous algorithms, the \pkg{greed} package also provides a
hierarchical approach to build a set of \(\ICL\) dominant solutions for
all the value of \(K\) between \(K^*\) and 1
\citep[see][for details]{come2021hierarchical}. This is achieved in the
same Bayesian paradigm with the \(\ICLex\) objective, now considering
the Dirichlet prior parameter \(\alpha\) as a regularization controlling
the granularity of the clustering and unlocking access to simpler,
coarser, solutions as it decreases towards \(0\). The method is based on
a log-linear approximation of \(\ICLex\) when \(\alpha\) is small, which
makes it computationally efficient, and produces a hierarchy of nested
partitions along with the sequence of the regularization parameters
which enabled the fusions
\(( \Clust^{(k)}, \,\alpha^{(k)} )_{k=K^\star,\ldots,1}\). Each of these
partitions is dominant in terms of \(\ICL\) when \(\alpha\) is in the
value range between \(\alpha^{(k-1)}\) and \(\alpha^{(k)}\). Such a
hierarchical processing allows for the exploration of the clustering at
coarser scales, together with an optimal ordering of the clusters which
is a powerful tool both for visual representation of the results and
their analysis. This final ordering is computed in an optimal fashion in
order to minimize the sum of merging costs between adjacent clusters
while respecting the order constraints imposed by the merge tree
\citep[Section 4]{come2021hierarchical}. This can be done efficiently
thanks to the dynamic programming algorithm of \citet{Bar2001}. It is
particularly useful for the dendrogram representation of the hierarchy
and other visualization tools described in \Cref{sec:applications}.

\hypertarget{implementation-details}{%
\subsection{Implementation details}\label{implementation-details}}

One of the main features of the \pkg{greed} package is its flexibility,
as it can handle any observational model, \textit{i.e.} any DLVM, for
which a tractable exact \(\ICL\) expression can be derived. The
implementation reflects this fact and, while most of the standard DLVMs
are implemented, new observational models can seamlessly be integrated
without impacting the main functions and API. Each observational model
must obey an interface and implements methods that compute the
difference of \(\ICLex\) induced by an elementary change of a solution
(\textit{i.e.} a swap or a merge) and to compute the observational part
of the \(\ICLex\) criterion in \Cref{eq:ICLex}, denoted as
\(\log \p(\Obs \mid \Clust, \bBeta)\). For the available models, the
swap and merge moves are efficiently implemented, only updating the
necessary terms rather than computing the whole \(\ICLex\) twice and
taking the differences. Moreover, whenever possible we took advantage of
possible computational speedups, \textit{e.g.} for Mixture of
regressions models we used Woodbury identity to efficiently perform
rank-one updates of the inverse of the covariance matrices induced by
the swaps.

The remaining part of the code base is generic for all models. On the
computational side, the main demanding methods were developed in
\proglang{C++} thanks to the \pkg{Rcpp} package \citep{Eddelbuettel2017}
taking advantages of sparse matrix computational efficiency provided by
the \pkg{RcppArmadillo} and \pkg{Matrix} packages
\citep{Eddelbuettel2014,Bates2019}. The greedy swap and merge methods
are implemented in \proglang{C++} and the high-level optimization
algorithms GA, hybrid GA, multiple restarts and seeded greedy algorithms
in \proglang{R}. Eventually, the \pkg{future} package
\citep{Bengtsson2019} was used to enable easy parallelization of the
computations in the hybrid GA and multistarts algorithms. Finally, the
observational model and the optimization algorithm are defined with S4
classes which we detail in the next section on some concrete use cases.

\paragraph{A note on numerical performances and algorithms complexities.}

\begin{CodeChunk}
\begin{figure}

{\centering \includegraphics[width=1\linewidth]{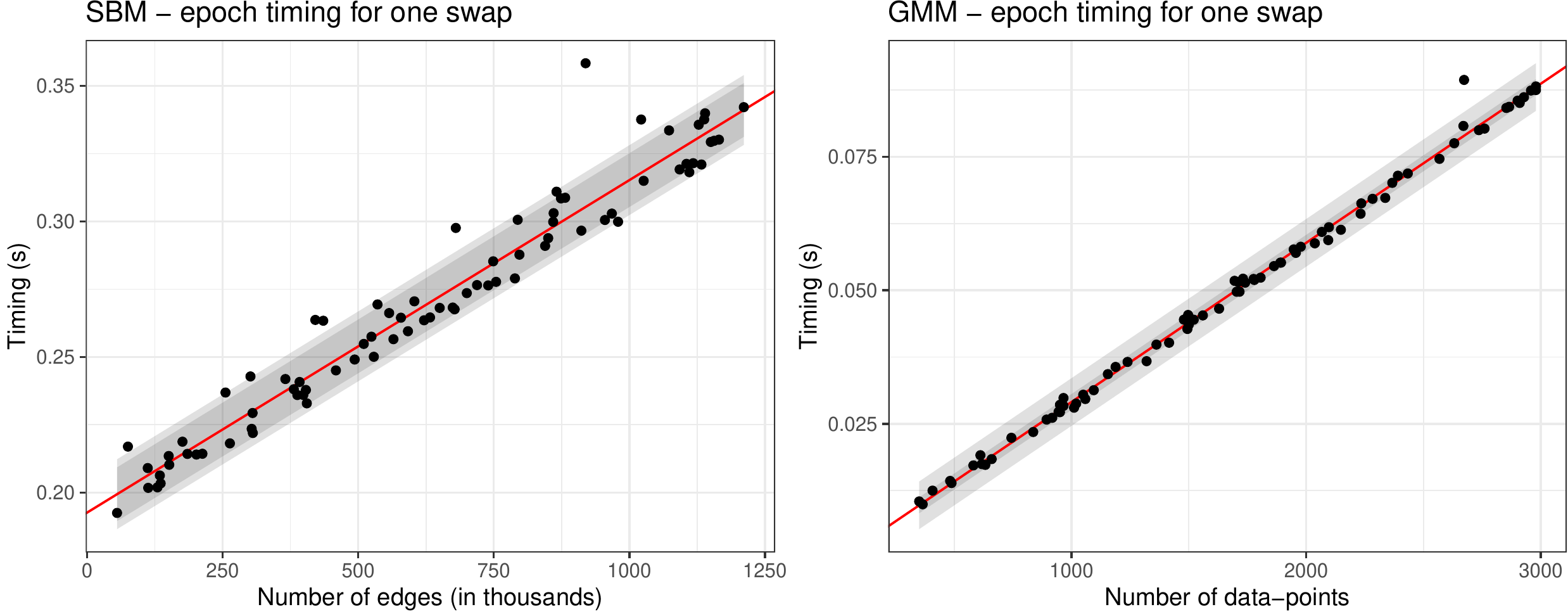} 

}

\caption[Timings of one swap pass for the SBM (left) and Diagonal-GMM (right) models]{Timings of one swap pass for the SBM (left) and Diagonal-GMM (right) models. For the SBM, the data were simulated using a planted partition model with 4 equiprobable clusters, 4000 nodes, an off-diagonal connection probability of $10^{-3}$ and a diagonal connection probability drawn at random between $0.01$ and $0.3$. This scheme leads to graphs with 250K to 1.25M edges. For the GMM, the datasets were simulated with a 3-component GMM in dimensions $10$. The number of data points was drawn at random between 300 and 3000. In both cases, the swap algorithm was initiated with $K=10$ clusters and 80 datasets were simulated. The timings were done on a Xeon(R) CPU, at 3.50GHz, and without parallelization.}\label{fig:plot-timing}
\end{figure}
\end{CodeChunk}

All the optimization algorithms proposed by the \pkg{greed} package rely
on combinations of locally optimal swap and merge moves. The time
complexity of the latter are model dependent: a whole epoch of swap
moves for an SBM (or it's degree corrected version) is in
\(\mathcal{O}(M\cdot K^2)\), with \(M\) the number of edges in the
graph\footnote{\pkg{greed} uses a sparse representation of the adjacency matrix to reach this complexity.}
and \(K\) the current number of clusters, whereas it reaches a
complexity in \(\mathcal{O}(N\cdot K \cdot D)\) for a Diagonal-GMM
model. Figure \ref{fig:plot-timing} clearly illustrates this
relationship on simulated data for both models. The timings are shown
for one epoch of a swap, and the linear relationships with the number of
edges for SBM and with the number of data-points for GMM clearly
appears. The timings are reasonable with respect to the problems sizes.

Regarding the merge moves, the complexities are also model dependent.
However, as in standard agglomerative procedures, the necessary
statistics only have to be computed once at the beginning. For example,
the cost of this operation is in \(\mathcal{O}(M)\) for the graphs
models and in \(\mathcal{O}(N \cdot P)\) for classical mixture, where
\(P\) is a model dependent constant. Then, the cost of computing the
\(\ICLex\) variation for \textit{one} merge is in \(\mathcal{O}(P)\),
with \(P=K\) for SBM like models and \(P=d\) for Diagonal-GMM models,
for example. Note that it does not depend on \(N\) or \(M\) anymore.
Merge moves are used twice in the algorithm: first in the GA in order to
merge redundant clusters after the cross-partition operator in
\Cref{eq:CrossPartition}, and second in the hierarchical clustering
algorithm used to complete the hierarchy after the genetic algorithm has
converged. These two use cases do not exhibit the same overall
complexity. Indeed, the GA takes advantage of the specific structure
yielded by the cross-partition operator to avoid exploring irrelevant
pairs of clusters. To do so, merge moves are only considered between
pairs of clusters that share a common parent in one of the two parent
partitions, see \citet[end of p.6]{come2021hierarchical} for details.
Moreover, the hierarchy is not built entirely during the GA greedy merge
steps, hence further reducing the computational costs. Thus, in
practice, the greedy merge heuristic does not reach its worst-case
complexity - which is in \(\mathcal{O}(K^2\cdot(K+P))\) - during the GA,
and it is quite economical compared to the swaps operations costs thanks
to the low values of \(K\). Concerning the final hierarchical step
algorithm, after the GA convergence, the complete hierarchy is built and
all possible merges are considered at each step. Therefore, its
complexity reaches the bound \(\mathcal{O}(K^2\cdot (K+P))\), which is
still reasonable for low values \(K=K^\star\) found by the GA.

Naturally, discussing each global algorithm complexity is quite
difficult, since they are influenced by many factors such as population
size, number of generations before convergence, mutations probabilities,
etc. Nevertheless, one may put in perspectives the different use cases
of the algorithms: the hybrid-GA and multistarts algorithms are quite
comparable in terms of computational time, provided that the population
size of the hybrid-GA is equal to the number of starting points of the
multistarts algorithm. The former being more efficient than the latter
on all the implemented models, it is the default of the package. The
seeded algorithm may be advantageous to users that want to quickly
explore large datasets, provided that the seeding algorithm is
efficient. Indeed, it is less computationally intensive since only one
swap epoch is performed, compared to several dozen for multistarts and
hybrid-GA.

A notable aspect of the package is its ability to handle graphs with
thousands of nodes in a very reasonable amount of time (less than 1
minute) making it competitive with respect to existing frequentist or
Bayesian approaches. As an illustration, in the simulation settings of
\Cref{fig:plot-timing}, the whole procedure runs in 30.2 seconds on
average for a SBM graph with 4000 nodes and more than 400K edges.

Finally, \pkg{greed} is seamlessly adaptable to parallel computing as
described in \Cref{subsec:future}. This can be easily used to obtain a
sensible speedup in performance when datasets become
large\footnote{For small datasets, the communication costs may exceed the gains from parallelism.}.

\hypertarget{guidelines-for-users}{%
\section{Guidelines for Users}\label{guidelines-for-users}}

\label{sec:guidelines}

This section provides an overview of a basic usage of the package. We
also describe the S4 classes and methods facilitating the manipulation
and exploration of the results. As a complement to this section and to
the usual \proglang{R} package documentation, a complete documentation
including model-specific vignettes is available as a \textbf{pkgdown}
website at \url{https://comeetie.github.io/greed/}.

\hypertarget{the-greed-function}{%
\subsection{The greed function}\label{the-greed-function}}

The package's main entry point is the \code{greed()} function, which
performs the clustering and presents a flexible API as illustrated in
the block of code below.

\begin{CodeChunk}
\begin{CodeInput}
R> library("greed")
R> sol = greed(
+    X                  , # dataset to cluster
+    K = 20             , # number of initial clusters (optional)
+    model = Gmm()      , # model to use and its prior parameters (optional)
+    alg = Hybrid()     , # algorithm to use and its parameters (optional)
+    verbose= FALSE    )  # verbosity level (optional)
\end{CodeInput}
\end{CodeChunk}

The \code{greed()} function only has one mandatory argument, \code{X},
which contains a dataset to cluster. All remaining arguments are
optional:

\begin{itemize}
  \item \code{K} represents the initial number of clusters and is set to 20 by default. Note that this value is an initial guess, and the available optimization algorithms may return a partition with more or less clusters.
  \item \code{model} is an S4 object inheriting of the abstract class \code{DlvmPrior} and containing the observational model used to compute $\ICLex$ along with its (model-specific) hyperparameters. If not provided, \code{greed} will try to infer a compatible model with the provided dataset. For example, if \code{X} is a symmetric binary sparse \code{dgCmatrix} the model will be set to an \code{Sbm()} model. Or else, if the provided dataset is a \code{data.frame} with only columns of \code{factor} type, an \code{Lca()} model will then be used. The list of available observational models, their prior parameters and allowed inputs is summarized in \Cref{tab:obsmodS4}. 
\item \code{alg} determines the clustering algorithm used by the \code{greed()} function. The algorithm is an S4 object inheriting from the abstract \code{Alg} class and containing the relevant hyperparameters (\textit{e.g.} probability of mutation, etc.). By default, the hybrid GA described in \Cref{sec:algo} is used, but other maximization heuristics may be used such as standard greedy hill-climbing with multiple random starts, or a standard GA without greedy local search. 
\end{itemize}

\hypertarget{analysing-the-clustering-result}{%
\subsection{Analysing the clustering
result}\label{analysing-the-clustering-result}}

The \code{greed()} function return an S4 object that inherits from the
\code{IclPath} S4 class, \textit{e.g.} a \code{SbmPath} object when the
model used is an \code{Sbm} or a \code{GmmPath} object when the model
used is a \code{Gmm}. Thus, any clustering result shares the same
structure and can be analyzed with the same set of S4 methods,
independently from the specified model, illustrating the generic aspect
of the methodology. These methods are listed in \Cref{tab:S4methods}
along with their description, and \Cref{sec:applications} presents
several uses in concrete examples.

\begin{table}[!ht]
    \renewcommand{\arraystretch}{1.2}
\begin{tabular}{lp{0.7\textwidth}}
\hline
\textbf{Method name} & \textbf{Description}  \\ \hline\hline
  \code{clustering(sol)} & Returns the estimated partition, integer encoded. \\ \hline
 \code{K(sol)} & Returns the number of clusters of the partition.  \\ \hline
 \code{ICL(sol)}  & Returns the ICL value of the partition. \\ \hline
\code{prior(sol)}  & Returns the \code{model} object used in the call to \code{greed}, with eventually the hyper-parameter values that were used during the fit and fixed automatically in a data-driven way. \\ \hline
\code{cut(sol, K)} & Cut the dendrogram at specified level \code{K} and returns the corresponding \code{IclPath} object. Similar to \code{stats::cutree()}. \\ \hline
\code{coef(sol)} & Returns the MAP estimates of the model parameters $\param$ conditionally on the estimated partition in a named list. \\ \hline
\code{plot(sol, type="tree")} & Returns a \pkg{ggplot2} object containing the desired plot \code{type}, \textit{e.g.} dendrogram of the hierarchy or any model-dependent representation of the clustering.  \\ \hline
\end{tabular}
\caption{\label{tab:S4methods}S4 methods associated to the \code{IclPath} class allowing analysing the results \code{sol} of the \code{greed()} function.}
\end{table}

As a quick illustration, we present the results and usage of these
methods in the case of graph clustering with SBM (this model will be
presented in more details in \Cref{subsec:sbm}) using a nested
hierarchical community structure in the simulation. The simulations
deals with 12 clusters arranged in three levels: 2 big clusters, each
composed of three medium clusters, themselves decomposed in two smaller
groups. We choose a community structure where nodes inside the same
blocks tend to connect more. The connectivity matrix \(\param\) of the
simulation is displayed in \Cref{fig:sbmsim} and the model in
\Cref{eq:sbm}.

\begin{CodeChunk}
\begin{figure}

{\centering \includegraphics[width=0.5\linewidth]{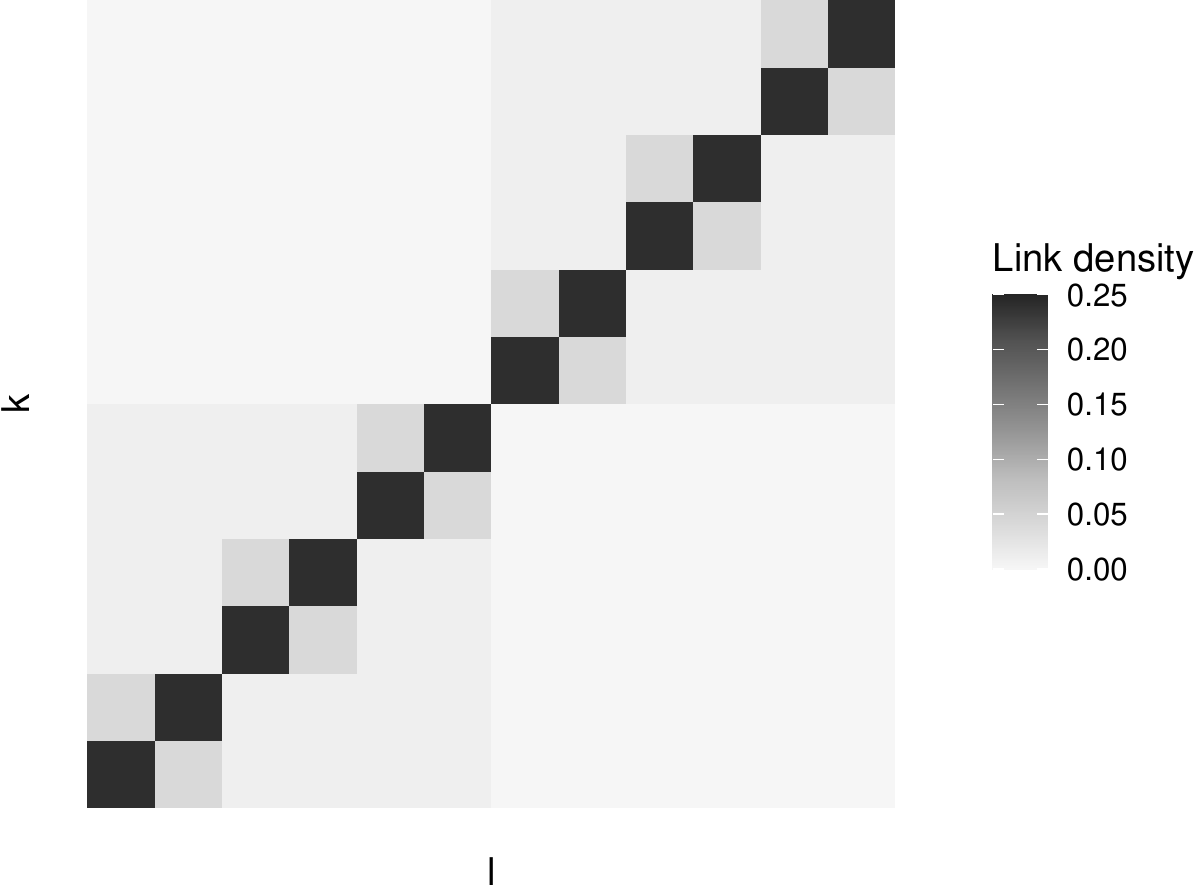} 

}

\caption[Connectivity matrix of the SBM hierarchical simulation]{Connectivity matrix of the SBM hierarchical simulation.}\label{fig:sbmsim}
\end{figure}
\end{CodeChunk}

\begin{CodeChunk}
\begin{CodeInput}
R> N  <- 800           # Number of node
R> K  <- 12            # Number of cluster
R> pi <- rep(1/K,K)    # Clusters proportions 
R> sbmsim <- rsbm(N,pi,theta) # Simulation
R> sol    <- greed(sbmsim$x, model = Sbm()) # Clustering
\end{CodeInput}
\end{CodeChunk}

The algorithm found the true clustering structure. In addition, the
dendrogram and the block adjacency matrix of \Cref{fig:example-dendo}
highlight the hierarchical structure of the dataset.

\begin{CodeChunk}
\begin{CodeInput}
R> tree   <- plot(sol, type="tree")
R> blocks <- plot(sol,type='blocks')
R> ggarrange(blocks,tree)
\end{CodeInput}
\begin{figure}

{\centering \includegraphics[width=0.9\linewidth]{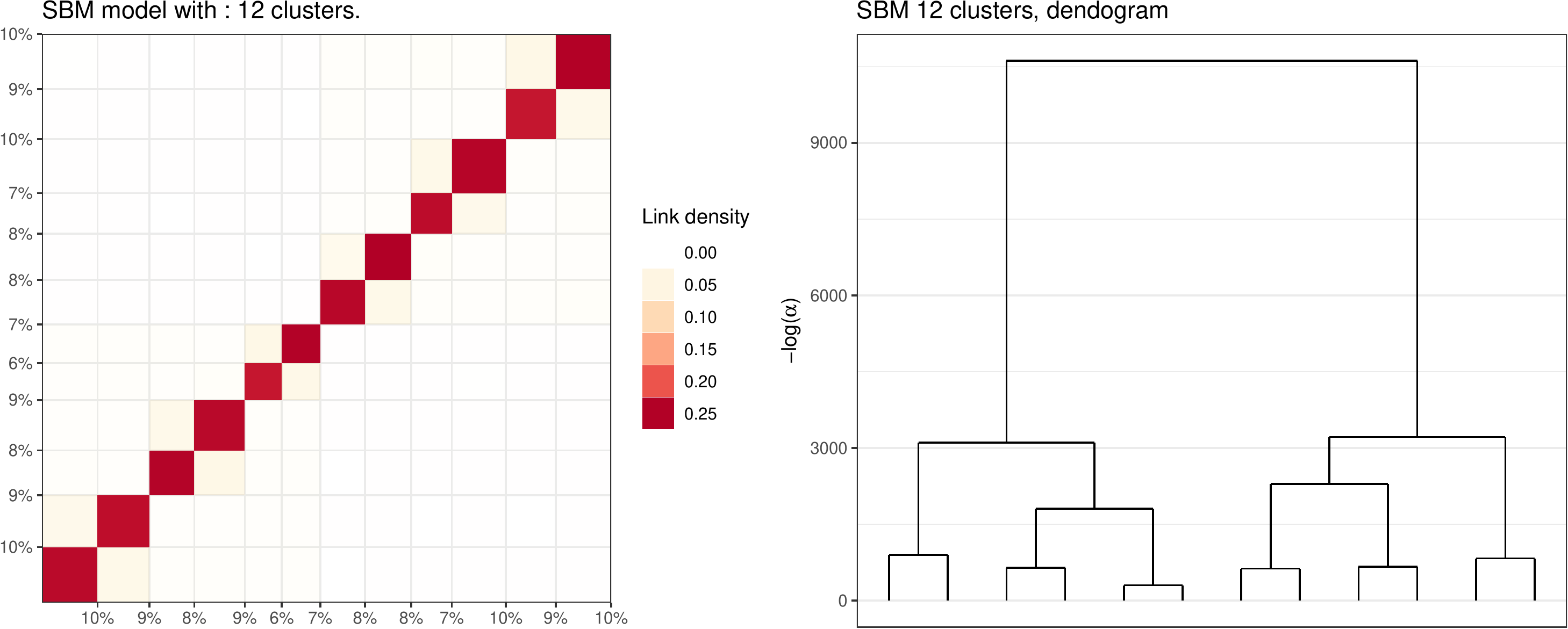} 

}

\caption[Blocks representation (left) and dendrogram (right) of the hierarchical SBM clustering found by greed]{Blocks representation (left) and dendrogram (right) of the hierarchical SBM clustering found by greed.}\label{fig:example-dendo}
\end{figure}
\end{CodeChunk}

The \code{cut()} function may then be used to inspect the clustering at
the ``medium'' (\(K=6\)) and ``top'' (\(K=2\)) levels of the hierarchy.
In the following code block, we illustrate the usage of the \code{K()},
\code{ICL()} and \code{clustering()} functions on these coarser
solutions.

\begin{CodeChunk}
\begin{CodeInput}
R> sol_medium = cut(sol, 6)
R> K(sol_medium)
\end{CodeInput}
\begin{CodeOutput}
[1] 6
\end{CodeOutput}
\begin{CodeInput}
R> ICL(sol_medium)
\end{CodeInput}
\begin{CodeOutput}
[1] -61080.25
\end{CodeOutput}
\begin{CodeInput}
R> sol_top = cut(sol, 2)
R> table(clustering(sol_top), sbmsim$cl)
\end{CodeInput}
\begin{CodeOutput}
   
     1  2  3  4  5  6  7  8  9 10 11 12
  1 54 52 73 78 71 63  0  0  0  0  0  0
  2  0  0  0  0  0  0 63 65 78 70 57 76
\end{CodeOutput}
\end{CodeChunk}

To finish, the \code{coef()} method may be used on \code{sol}, returning
an estimate \(\hat{\param}\) of the connectivity matrix. Note how, in
this specific example, the cluster permutation given by the dendrogram
allows to directly compare \(\hat{\param}\) to the true \(\param\),
which is typically not the case in regular inference procedure.

\begin{CodeChunk}
\begin{CodeInput}
R> hat_theta= coef(sol)$theta
R> mean(abs(hat_theta-theta)) # l1 relative error, no permutation needed.
\end{CodeInput}
\begin{CodeOutput}
[1] 0.001154612
\end{CodeOutput}
\end{CodeChunk}

\hypertarget{available-models-and-prior-specification}{%
\subsection{Available models and prior
specification}\label{available-models-and-prior-specification}}

\label{subsec:available-models} The list of currently implemented DLVM
is available in \Cref{tab:obsmodS4}, describing the name of the
hyperparameters as well as the allowed inputs \code{X} for each model. A
summary of implemented models is also available through the helper
function \code{available_models()}.

\begin{table}[!ht]
    \renewcommand{\arraystretch}{1.2}
\begin{tabular}{llL{0.25\textwidth}L{0.25\textwidth}}
        \hline
          S4 class name & Data type & Prior parameters & Allowed input\\
 \hline
 \hline

\code{MoM()}  & Counts & \code{alpha}, \code{beta} &  \code{matrix}, \code{dgCmatrix}, \code{data.frame} \\\hline

\code{Lca()} & Factors & \code{alpha}, \code{beta} &   \code{data.frame} \\\hline

\code{Sbm()} & Graphs & \code{alpha}, \code{a0}, \code{b0}& \code{matrix}, \code{dgCmatrix}, \code{igraph}\\\hline
 \code{DcSbm()} & Graphs & \code{alpha}, \code{p}$^{(*)}$&\code{matrix}, \code{dgCmatrix}, \code{igraph}\\\hline
\code{DcLbm()}& Bipartite graph & \code{alpha}, \code{p}$^{(*)}$&\code{matrix}, \code{dgCmatrix}\\\hline

\code{Gmm()} & Continuous & \code{alpha}, \code{tau}, \code{N0}$^{(*)}$, \code{epsilon}$^{(*)}$, \code{mu}$^{(*)}$ &  \code{matrix}, \code{data.frame} \\\hline

 \code{DiagGmm()} & Continuous & \code{alpha}, \code{tau}, \code{kappa}, \code{mu}$^{(*)}$, \code{beta}$^{(*)}$ &  \code{matrix}, \code{data.frame} \\\hline

\code{MoR(y~x)}  & Continuous & \code{alpha}, \code{tau}, \code{N0}$^{(*)}$, \code{espilon}$^{(*)}$, \code{formula} & \code{data.frame} \\\hline
\code{CombinedModels(models)} & Mixed & - & \code{list}
\end{tabular}
\caption{\label{tab:obsmodS4}S4 classes of the observational models handled by the \pkg{greed} package along with their prior hyperparameters name and the classes of the input datasets that these models can process. Informative hyperparameters fixed in a data-driven way are highlighted with a $^{(*)}$ symbol.}
\end{table}

Any observational model may be created with prior hyperparameters
\(\bBeta\) values automatically chosen, \textit{e.g.} by just calling
\code{Gmm()} or \code{Sbm()}. In this case, the chosen value for the
prior hyperparameters corresponds to non-informative uniform prior
whenever possible, as in the MoM, LCA an SBM models. Whenever
non-informative priors are unavailable, they are chosen in a data-driven
way. For example, the \code{p} hyperparameter for \code{DcSbm} and
\code{DcLbm} is set to the average connection probability between two
nodes. The help and documentation of each model precisely describes its
hyperparameters and their default values. Moreover, if any prior
information is available, their value may be specified by the user, one
may for example use \code{Gmm(tau=0.3)} to create a \code{Gmm} model
prior with hyperparameter \code{tau} (controlling the variance of the
Gaussian prior on the means) equal to \(0.3\). However, two models have
one mandatory argument the Mixture of Regression model must be provided
with a formula describing the linear model to use \code{MoR(y~x)} and
eventually, the \code{CombinedModels} class that enable the stacking of
several observational models on the same individuals, must be provided
with a list of models to use for each view, its usage will be covered in
\Cref{subsec:combined}.

\hypertarget{controlling-optimization-algorithms-and-their-hyperparameters}{%
\subsection{Controlling optimization algorithms and their
hyperparameters}\label{controlling-optimization-algorithms-and-their-hyperparameters}}

As explained above, optimization algorithms used by \code{greed()} are
represented as S4 object inheriting from the class \code{Alg}. In this
representation, an algorithm has all its hyperparameters stored as
attributes of the object, allowing fine-grained control when
instantiating the object. For example, in the case of the hybrid GA
briefly introduced in \Cref{sec:algo} - and used as a default by
\code{greed()} - one may customize the behaviour via its \code{Hybrid()}
constructor and specify some of the following parameters:

\begin{itemize}
\item \code{pop_size}: size of the populations (default to 20)
\item \code{nb_max_gen}: maximal number of generations (default to 10)
\item \code{prob_mutation}: mutation probability (default to 0.25)
\item \code{Kmax}: maximum number of clusters (default to 100)
\end{itemize}

A user desiring to increase the exploration capacities of the algorithm
may increase the population size, at the expanse of an increased
computational time, by using something like \code{Hybrid(pop_size=50)}.
Increasing the two other parameters \code{nb_max_gen} and
\code{prob_mutation} will also help with the exploration capacities of
the algorithm but to a lesser extent. Eventually, the \code{Kmax}
parameter allows the user to set an upper bound for the desired number
of cluster. Indeed, this specific algorithm may in fact find a larger
number of clusters than the initial value provided to \code{greed()}.
Thus, this upper bound avoids the computations of solutions with too
many clusters.

The list of implemented algorithm is available via the helper function
\code{available_algorithms()}. \code{Seed()} do not have any
hyperparameters while \code{Multistarts()} only one - \code{nb_start} -
corresponding to the number of different random initialization used.
Eventually, \code{Genetic()} share the same parameters as
\code{Hybrid()}.

\hypertarget{parallel-computing.}{%
\subsection{Parallel computing.}\label{parallel-computing.}}

\label{subsec:future}

Parallel processing is implemented in \pkg{greed} through the
\pkg{future} package, providing a unified and straightforward framework
for sequential and parallel processing. A \code{future::plan} must
simply be set by the user (default is sequential with no parallelism).
For instance, in order to run \code{greed()} on 10 cores in multiple
sessions, the following commands should be used before any call to
\code{greed()}.

\begin{CodeChunk}
\begin{CodeInput}
R> future::plan(multisession,workers=10)
\end{CodeInput}
\end{CodeChunk}

Finally, note that \pkg{future} supports many parallel processing
paradigms and options. Details are provided in \code{plan()}
documentation as well as in the package's vignettes. In \pkg{greed}, the
parallelism will be used with the \code{Multistarts()} (one process per
stating points), \code{Genetic()} and \code{Hybrid()} algorithms (one
process per elements of the population). The interest of enabling
parallel processing can be limited and even counterproductive for small
datasets so it is best to reserve its use for massive datasets.

\hypertarget{applications}{%
\section{Applications}\label{applications}}

\label{sec:applications}

In this section, we demonstrate the use of the \pkg{greed} package on
continuous, categorical, graph and heterogeneous data clustering
problems through the lens of illustrative real datasets. Each of these
applications uses some underlying standard observational model,
\textit{i.e.} a DLVM, which we will briefly introduce, along with the
model-specific hyperparameters \(\bBeta\).

\hypertarget{continuous-data-clustering-with-gaussian-mixtures}{%
\subsection{Continuous data clustering with Gaussian
mixtures}\label{continuous-data-clustering-with-gaussian-mixtures}}

We are interested in clustering a set of \(n\) multivariate observation
in dimension \(p\), \(\Obs = \{ \obs_i \}_{i=1, \ldots, n}\). As
explained in \Cref{sec:theory}, Gaussian mixture models (GMMs) are a
widely used DLVM in this context. Without any constraints, the Bayesian
formulation of GMMs leading to a tractable exact ICL expression uses a
Normal and inverse-Wishart conjugate prior on the mean and covariances
\(\param = (\bm{\mu}_k, \bm{\Sigma}_k)_k\) \citep{Bertoletti2015}. This
prior is defined with hyperparameters
\(\bBeta = (\bm{\mu}, \tau, n_0, \bm{\varepsilon})\) and the
hierarchical formulation is as follows:

\begin{equation}
\label{eq:gmm}
\begin{aligned}
\pi&\sim \textrm{Dirichlet}_K(\alpha)\\
Z_i&\sim \mathcal{M}(1,\pi)\\
\bm{\Sigma}_k^{-1} & \sim \textrm{Wishart}(\bm{\varepsilon}^{-1},n_0)\\
\bm{\mu}_k&\sim \mathcal{N}(\bm{\mu},\frac{1}{\tau} \bm{\Sigma}_k)\\
X_{i}|Z_{ik}=1 &\sim \mathcal{N}(\bm{\mu}_k, \bm{\Sigma}_{k})\\
\end{aligned}
\end{equation}

These priors are informative and may therefore have a sensible impact on
the obtained results. By default, the priors parameters are set in a
data driven fashion as follows: \(\mu=\bar{\Obs}\), \(\tau=0.01\),
\(n_0=d\), \(\varepsilon=0.1\textnormal{diag}(\hat{\bm{\Sigma}}_\Obs)\),
where \(\widehat{\bm{\Sigma}}_\Obs\) denotes the empirical covariance
matrix. These values were chosen to accommodate a variety of situation
but they can still be specified by the user and their impact
investigated through a sensitivity analysis.

When the number of variables \(p\) becomes large, it can be of interest
to reduce the flexibility of a GMM by adding a specific set of
constraints. The diagonal covariance model, with constraint
\(\bm{\Sigma}_k = \textnormal{diag}(\sigma_{k1}, \ldots, \sigma_{kp})\),
is implemented in \pkg{greed} through its \code{DiagGmm} S4 class. Its
hierarchical formulation is similar to that of \Cref{eq:gmm}, except the
Wishart prior on \(\bm{\Sigma}_k\) is now replaced by a Gamma prior on
each \(\sigma_{kj}\) with shape \(\kappa\) and rate \(\beta\)
hyperparameters. The latter have default values set to \(1\) and to the
mean of the empirical columns variances respectively.

\hypertarget{the-diabetes-data}{%
\subsubsection{The diabetes data}\label{the-diabetes-data}}

Let us describe a first use case on the \code{diabetes} dataset from the
\pkg{mclust} package a small dataset suitable for illustration purposes.
The data describes \(p=3\) biological variables for \(n=145\) patients
with \(K=3\) different types of diabetes: normal, overt and chemical. We
are interested in the clustering of this dataset to see if the three
biological variables can discriminate between the three diabetes type.
We begin by applying the \code{greed} function with a \code{Gmm} object
with default hyperparameters, and compare the obtained clusters with the
known diabetes types of the patients.

\begin{CodeChunk}
\begin{CodeInput}
R> data("diabetes")
R> X=diabetes[,-1]
R> sol = greed(X,model=Gmm())
\end{CodeInput}
\end{CodeChunk}

\begin{CodeChunk}
\begin{CodeInput}
R> table(diabetes$class, clustering(sol)) 
\end{CodeInput}
\begin{CodeOutput}
          
            1  2  3  4
  Chemical 11 24  1  0
  Normal   74  2  0  0
  Overt     0  5  0 28
\end{CodeOutput}
\end{CodeChunk}

The \code{gmmpairs} function allows to get an overview of the clustering
results by plotting every biplot combinations of the original variables
with the estimated clusters membership and Gaussian ellipses highlighted
in each cluster.

\begin{CodeChunk}
\begin{CodeInput}
R> gmmpairs(sol, X)
\end{CodeInput}


\begin{center}\includegraphics[width=0.9\linewidth]{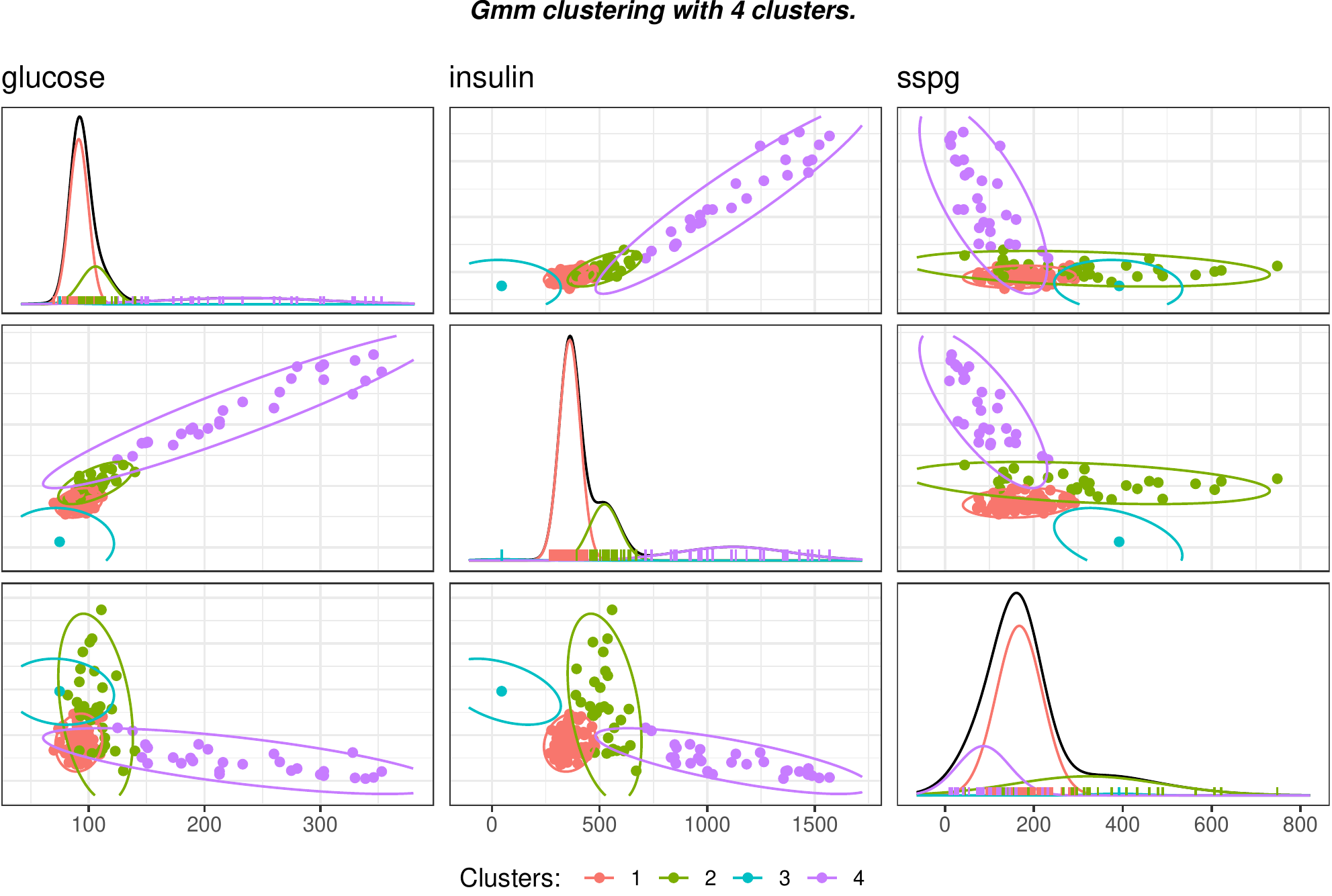} \end{center}

\end{CodeChunk}

If a user wants to experiment with other values of the prior
hyperparameters, the most important is \(\mathbf{\varepsilon}\), which
control the clusters' covariance matrices Wishart prior. This amounts to
specify the prior on the dispersion inside each class. For instance, one
may specify a priori belief that the variance is small inside clusters,
which amounts to diminish the \(0.1\) coefficient in front of
\(\hat{\mathbf{\Sigma}}_{\mathbf{X}}\). In this case, it makes sense to
decrease \(\tau\) in the same proportions to keep a flat priors on the
clusters means. For the diabetes data such a choice leads to an
interesting solution, where the strong prior constraint leads to one
cluster being created to fit one outlier in the ``chemical'' diabetes.

\begin{CodeChunk}
\begin{CodeInput}
R> sol_dense = greed(X,model=Gmm(epsilon=0.01*diag(diag(cov(X))), tau =0.001))
\end{CodeInput}
\end{CodeChunk}

\begin{CodeChunk}
\begin{CodeInput}
R> gmmpairs(sol_dense,X)
\end{CodeInput}
\begin{figure}

{\centering \includegraphics[width=0.9\linewidth]{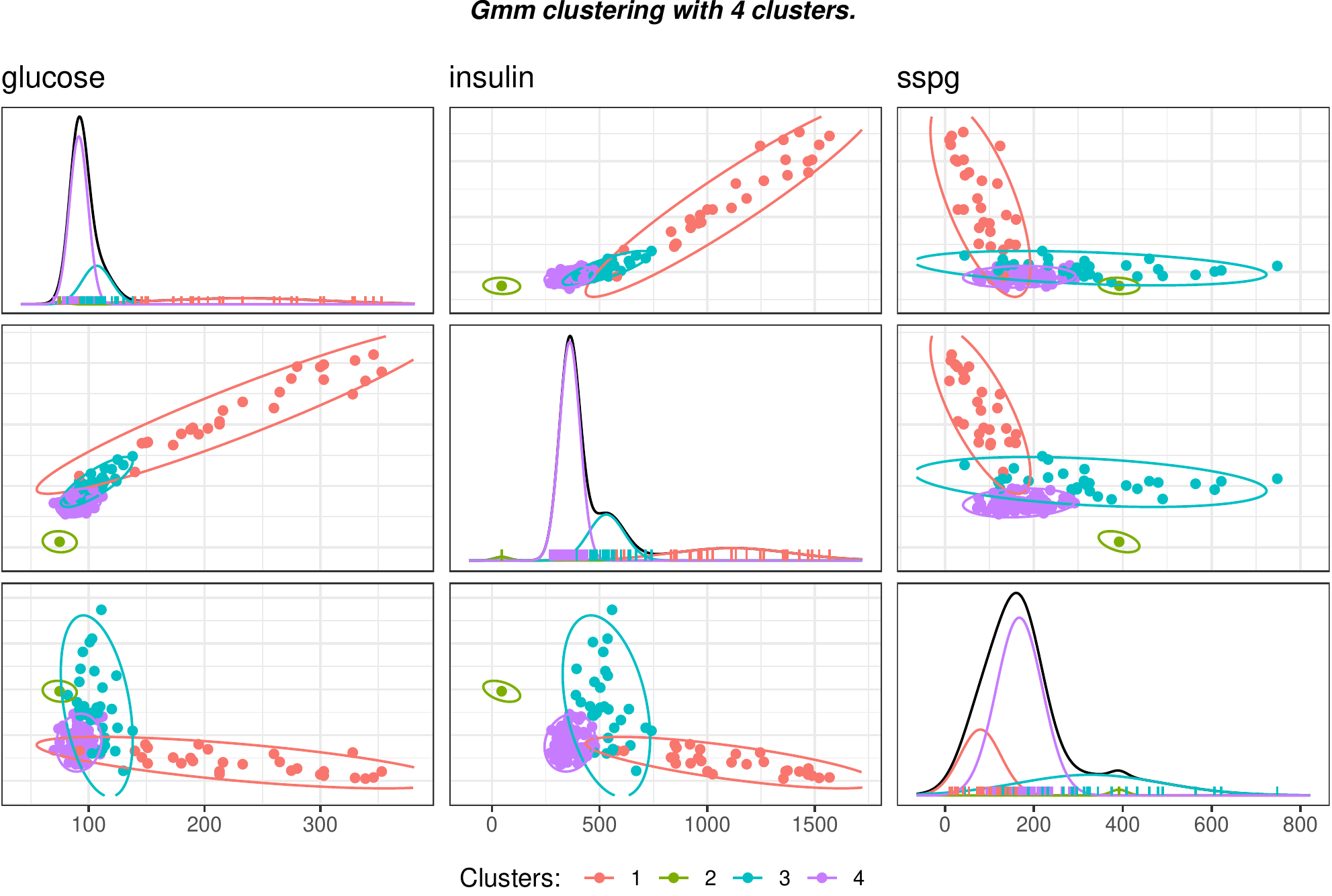} 

}

\caption[Matrix pairs plots of the clustering with user specified hyperparameters on the diabetes data]{Matrix pairs plots of the clustering with user specified hyperparameters on the diabetes data.}\label{fig:pairplot-dense}
\end{figure}
\end{CodeChunk}

\hypertarget{categorical-data-clustering-with-latent-class-analysis}{%
\subsection{Categorical data clustering with Latent class
analysis}\label{categorical-data-clustering-with-latent-class-analysis}}

\hypertarget{the-model}{%
\subsubsection{The model}\label{the-model}}

We are interested in the clustering of categorical datasets, which are
typically found in survey data or item response theory. In this context,
we observe \(n\) individuals described by \(p\) variables, taking one
among \(d_j\) modalities for each variable \(j\). Latent class analysis
(LCA) is a generative model for categorical data clustering which posits
conditional independance of the factor variables conditionally on the
(unknown) partition, hence fitting the definition of a DLVM. Below is a
description of its Bayesian formulation with the use of proper conjugate
priors

\begin{equation}
\begin{aligned}
\pi&\sim \textrm{Dirichlet}_K(\alpha),\\
\forall k, \forall j, \quad \theta_{kj} &\sim \textrm{Dirichlet}_{d_j}(\beta), \\
Z_i&\sim \mathcal{M}_K(1,\pi),\\
\forall j=1, \ldots, p, \quad X_{ij}|Z_{ik}=1 &\sim \mathcal{M}_{d_j}(1, \theta_{kj}),\\
\end{aligned}
\end{equation}

For each cluster \(k\) and variable \(j\), the vector \(\theta_{kj}\)
represents the probability of each of the \(d_j\) modalities. With the
choice of priors above, the LCA model admits an exact ICL expression
similar to the mixture of multinomials
model\footnote{The derivation may be found in Section 3 of the supplementary materials of \citet{come2021hierarchical}}.
Specifying priors hyper-parameters for such a model is less sensible
than in the GMM case since uninformative priors are available when there
is a lack of prior information.

\hypertarget{illustration-the-mushroom-dataset}{%
\subsubsection{Illustration: the mushroom
dataset}\label{illustration-the-mushroom-dataset}}

This dataset originally comes from UCI Machine Learning Repository and
describes 8124 mushrooms with 22 phenotype categorical variables. Each
mushroom is classified as ``edible'' or ``poisonous'' and the goal is to
recover the mushroom class from its phenotype. The \textbf{mushroom}
dataset is attached to the \pkg{greed} package and can be loaded as
follow

\begin{CodeChunk}
\begin{CodeInput}
R> data("mushroom")
\end{CodeInput}
\end{CodeChunk}

We illustrate the method on a subsample of size \(n=1000\) mushrooms.

\begin{CodeChunk}
\begin{CodeInput}
R> X <- mushroom[,-1]
R> subset   <- sample(1:nrow(X), size = 1000)
R> labels   <- mushroom$edibility[subset]
R> sol_mush <- greed(X[subset,], model=Lca())
\end{CodeInput}
\end{CodeChunk}

The hybrid genetic algorithm found a solution with \(K=13\) clusters
which is quite over-segmented. However, the hierarchical structure of
the clustering depicted in Figure \ref{fig:dendo-mushroom} (and
generated with the code block below) displays a good separation between
two main types of clusters.

\begin{CodeChunk}
\begin{CodeInput}
R> plot(sol_mush, type='tree')
\end{CodeInput}
\begin{figure}

{\centering \includegraphics[width=0.5\linewidth]{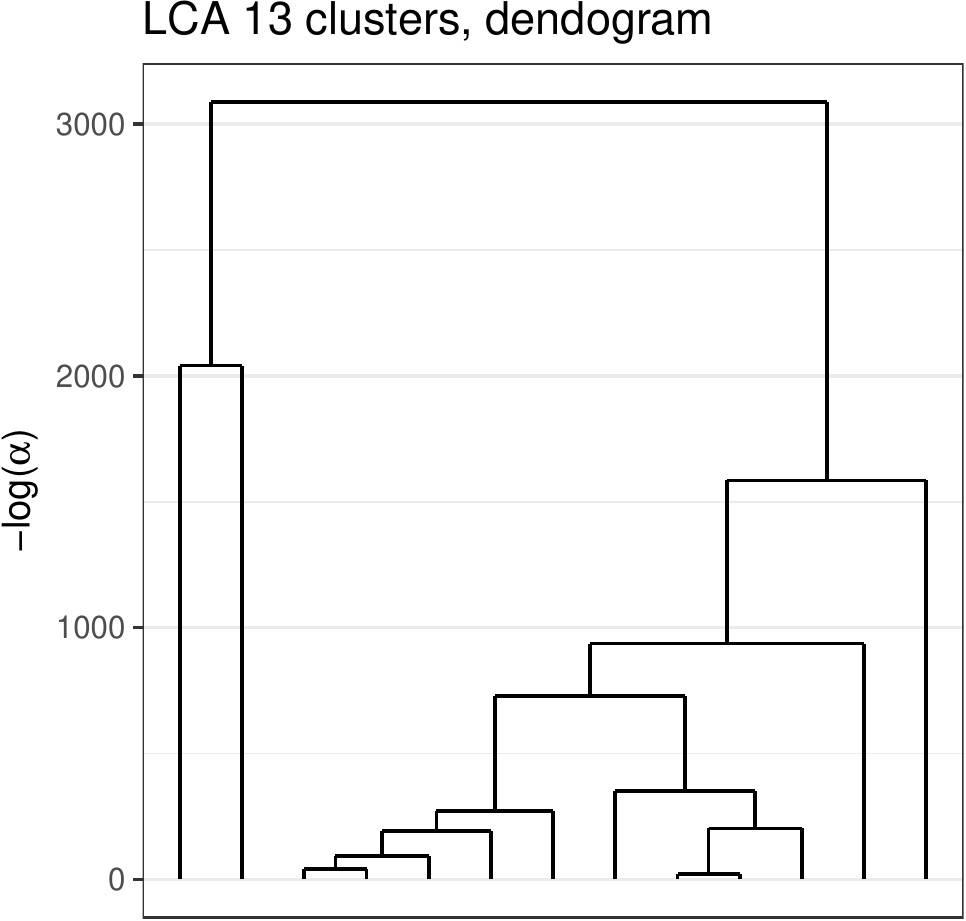} 

}

\caption[Dendrogram obtained by greed with the Lca model on the Mushroom dataset]{Dendrogram obtained by greed with the Lca model on the Mushroom dataset.}\label{fig:dendo-mushroom}
\end{figure}
\end{CodeChunk}

A clear hierarchical structure appears, with \(K=2\) or \(K=4\) main
clusters. Thus, we can cut the tree at this height and look at the
solution.

\begin{CodeChunk}
\begin{CodeInput}
R> sol4 = cut(sol_mush, 4)
R> table(labels, clustering(sol4)) 
\end{CodeInput}
\begin{CodeOutput}
      
labels   1   2   3   4
     e   0   0 315 208
     p 206 159 112   0
\end{CodeOutput}
\end{CodeChunk}

Here, we clearly see that the order of merges is consistent with the
edibility labels. While some poisonous mushrooms have been categorized
as edible, this might be the consequence of the way the labels have been
set, since mushrooms for which the edibility status was \emph{unknown}
were classified as \emph{poisonous} by default. While this choice is
reasonable from a strict health perspective, it might be too
conservative. Furthermore, as the data documentation specifies,
'`\textit{there is no simple rule for determining the edibility of a mushroom}''.
Thus, the unsupervised problem is hard and the obtained clustering is
already satisfactory.

\hypertarget{graph-clustering-with-the-stochastic-block-model}{%
\subsection{Graph clustering with the stochastic block
model}\label{graph-clustering-with-the-stochastic-block-model}}

\label{subsec:sbm}

\hypertarget{the-model-1}{%
\subsubsection{The model}\label{the-model-1}}

Graph data arise in various scientific fields from biology to sociology,
and accounts for relationship between objects. These objects are
expressed as \emph{nodes}, while a relationship between two objects is
expressed as an \emph{edge}. Hence, graph data may be expressed and
stored in an \emph{adjacency matrix} \(\Obs = \{ x_{ij} \}\) where
\(x_{ij}=1\) means that objects \(i\) and \(j\) are connected.

Stochastic block model (SBMs) form a family of random graph models for
the adjacency matrix \(\mathbf{X}\), widely used for graph clustering.
In these models, the probability of an edge \((i,j)\) is driven by the
cluster membership of node \(i\) and \(j\), hence the \textbf{block}
terminology.

They can be expressed in the DLVM framework and the \textbf{greed}
package handles two of these models: the classical binary SBM and its
degree-corrected variant with Poisson emissions. The Bayesian
formulation of a binary SBM is as follows

\begin{equation}
\label{eq:sbm}
\begin{aligned}
\pi&\sim \textrm{Dirichlet}_K(\alpha),\\
\theta_{k,l} & \sim \textrm{Beta}(a_0, b_0), \\
Z_i&\sim \mathcal{M}(1,\pi),\\
\forall (i,j), \quad x_{ij} \mid Z_{ik}Z_{jl}=1& \sim \mathcal{B}(\theta_{k,l}).
\end{aligned}
\end{equation}

This model class is implemented in the \code{Sbm} class. Here, the model
hyperparameters are:

\begin{itemize}
\tightlist
\item
  \(\alpha\) which is set to \(1\) by default.
\item
  The beta distribution parameters \(a_0\) and \(b_0\) on the
  connectivity matrix \(\mathbf{\theta}\). A non-informative prior can
  be chose with \(a_0=b_0=1\), which is the default value in \code{Sbm}.
\end{itemize}

Note that the \pkg{greed} package also handles the degree-corrected
variant of SBM in the \code{?DcSbm} class, allowing for integer valued
edges. The underlying model and its DLVM formulation are described in
depth in the Supplementary Materials of \citet{come2021hierarchical}.

\hypertarget{a-real-data-example-the-book-dataset}{%
\subsubsection{A real-data example: the Book
dataset}\label{a-real-data-example-the-book-dataset}}

The \textbf{Books} dataset was gathered by Valdis Krebs and is attached
to the \pkg{greed} package. It consist of an undirected co-purchasing
network of \(N=105\) books on US politics. Two books have an edge
between them if they have been frequently co-purchased together. We have
access to the labels of each book according to its political
inclination: conservative (``n''), liberal (``l'') or neutral (``n'').
We will compare the fit obtained with \code{greed()} and the two network
models namely \code{Sbm()} and
\code{DcSbm()}.\footnote{We emphasize that the \code{greed()} function is able to automatically detect if the network model type should be "directed" or "undirected". We specified it in the code chunk for the sake of clarity.}

\begin{CodeChunk}
\begin{CodeInput}
R> data("Books")
R> sol_dcsbm <- greed(Books$X,model = DcSbm(type="undirected"))
R> ICL(sol_dcsbm)
\end{CodeInput}
\begin{CodeOutput}
[1] -1344.987
\end{CodeOutput}
\begin{CodeInput}
R> sol_sbm   <- greed(Books$X,model = Sbm(type="undirected"))
R> ICL(sol_sbm)
\end{CodeInput}
\begin{CodeOutput}
[1] -1252.718
\end{CodeOutput}
\end{CodeChunk}

For this dataset, the regular SBM model seems to reach a better ICL
solution than its degree-corrected variant. Still,
\Cref{fig:blocks-book} allows visualizing both aggregated adjacency
matrices and comparing them.

\begin{CodeChunk}
\begin{CodeInput}
R> bl_sbm = plot(sol_sbm,type='blocks')
R> bl_dcsbm = plot(sol_dcsbm,type='blocks')
R> ggarrange(bl_sbm,bl_dcsbm)
\end{CodeInput}
\begin{figure}

{\centering \includegraphics[width=1\linewidth]{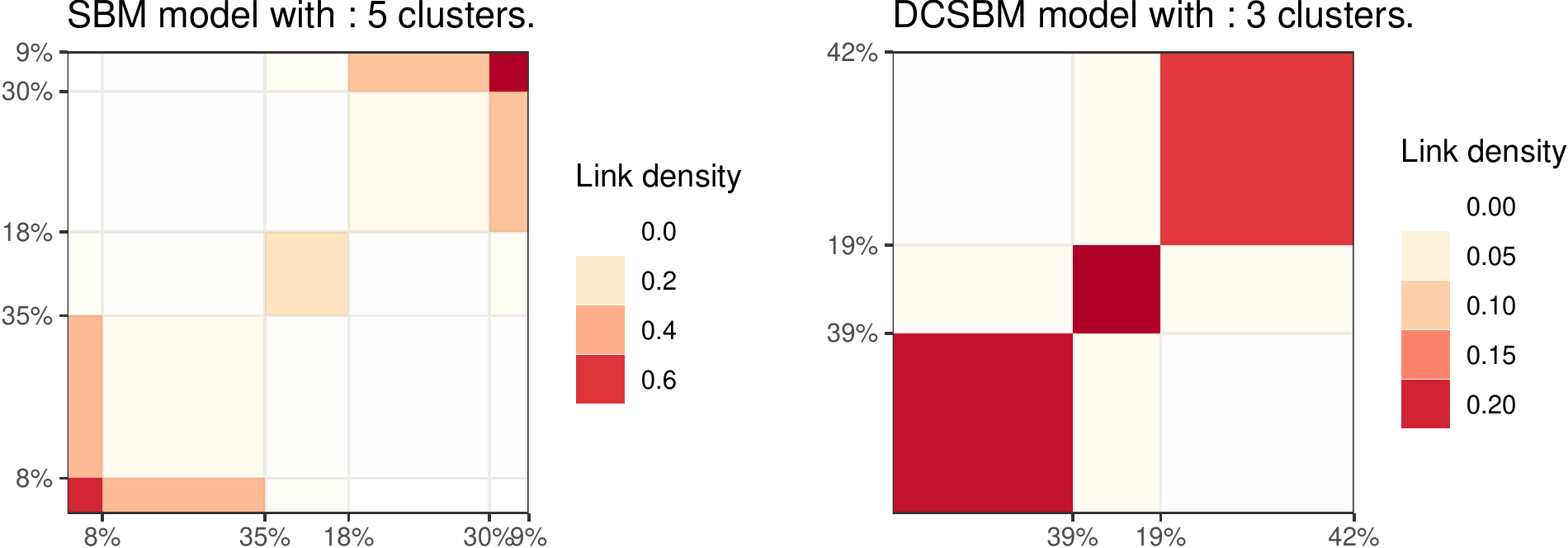} 

}

\caption[Block matrix representation of the DcSbm and Sbm solution found with greed on the Book network]{Block matrix representation of the DcSbm and Sbm solution found with greed on the Book network.}\label{fig:blocks-book}
\end{figure}
\end{CodeChunk}

In addition, external \proglang{R} packages may be used to display a
graph layout with node color as clusters and node size as book
popularity (computed using centrality degree). In
\Cref{fig:book-layout}, we represent the result for the SBM solution
with 5 clusters. One can see a hierarchical clustering structure
appearing, with a central cluster of neutral books in between two
densely connected set. In each of these two dense set, there is a clear
distinction between popular books (heavily purchased) and more
peripheric ones, indicated by node size.

\begin{CodeChunk}
\begin{CodeInput}
R> library("ggraph")
R> library("tidygraph")
R> library("igraph")
R> 
R> graph <- igraph::graph_from_adjacency_matrix(Books$X) %>% 
+   as_tbl_graph() %>% 
+   mutate(Popularity = centrality_degree())  %>% 
+   activate(nodes) %>%
+   mutate(cluster=factor(clustering(sol_sbm),1:K(sol_sbm)))
R> 
R> # plot using ggraph
R> ggraph::ggraph(graph, layout = 'kk') + 
+   geom_edge_link() + 
+   geom_node_point(aes(size = Popularity,color=cluster)) +
+   theme_bw()
\end{CodeInput}
\begin{figure}

{\centering \includegraphics[width=0.7\linewidth]{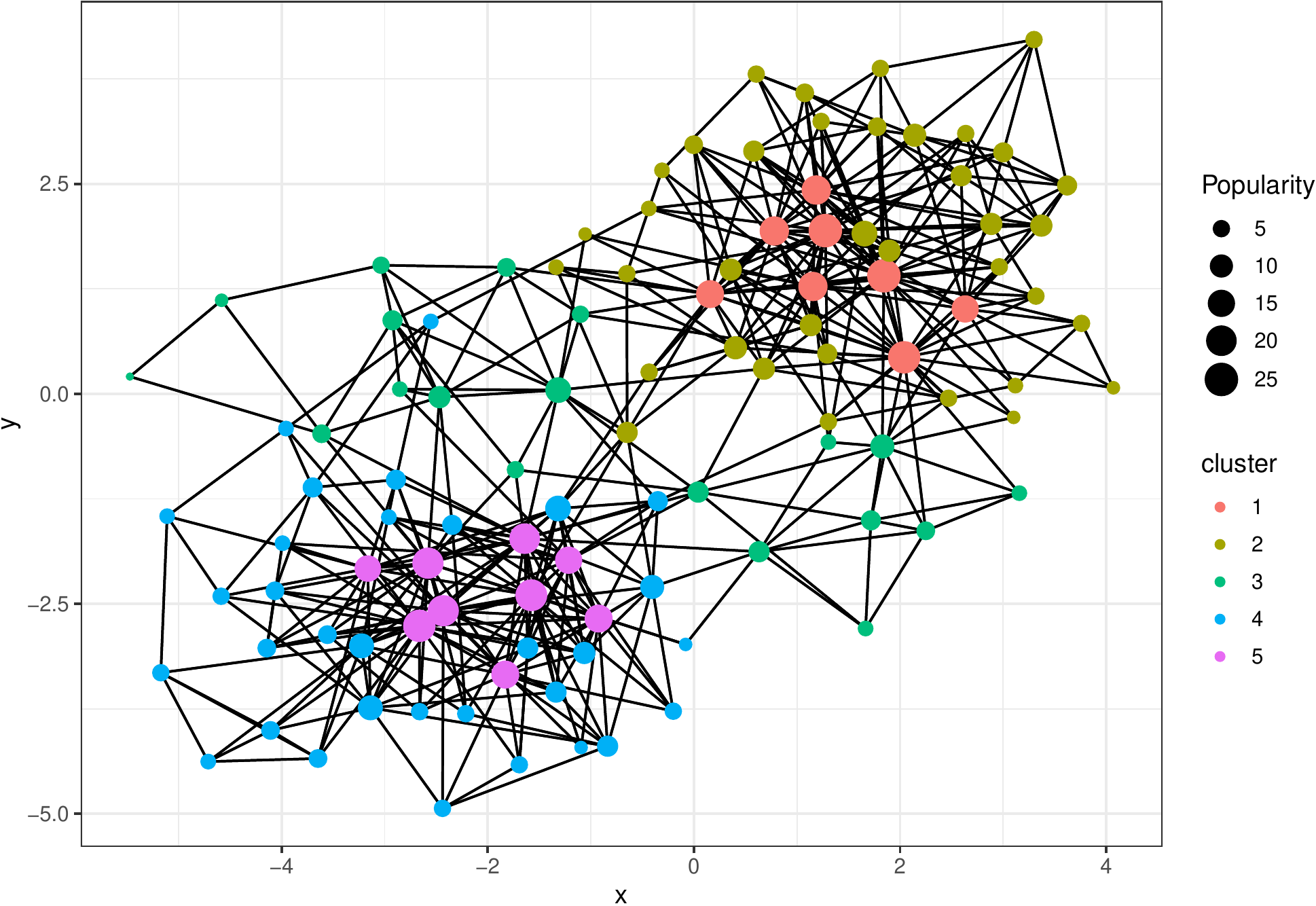} 

}

\caption[Ggraph plot of the book network with node colors indicating the clustering found by greed with an Sbm model]{Ggraph plot of the book network with node colors indicating the clustering found by greed with an Sbm model.}\label{fig:book-layout}
\end{figure}
\end{CodeChunk}

Finally, we can look at both models solutions and their confusion
matrix. We see that both partitions make sense according to the
available political labels.

\begin{CodeChunk}
\begin{CodeInput}
R> # Regular SBM
R> table(clustering(sol_sbm), Books$label)
\end{CodeInput}
\begin{CodeOutput}
   
     c  l  n
  1  8  0  0
  2 34  0  3
  3  6  5  8
  4  1 29  2
  5  0  9  0
\end{CodeOutput}
\begin{CodeInput}
R> # DcSbm
R> table(clustering(sol_dcsbm), Books$label)
\end{CodeInput}
\begin{CodeOutput}
   
     c  l  n
  1  1 38  2
  2  7  5  8
  3 41  0  3
\end{CodeOutput}
\end{CodeChunk}

\hypertarget{mixed-type-data-clustering-with-combined-models}{%
\subsection{Mixed-type data clustering with combined
models}\label{mixed-type-data-clustering-with-combined-models}}

\label{subsec:combined}

One of the distinctive features of the DLVM framework is that several
observational models may be combined together. This is particularly
useful when dealing with heterogeneous or mixed-type data, frequently
arising in modern applications.

\hypertarget{statistical-justification}{%
\subsubsection{Statistical
justification}\label{statistical-justification}}

In this context, we have different representations - or \emph{views} -
related to the same \(n\) objects,
\(\Obs = \left\{\Obs_v\right\}_{v=1, \ldots, V}\) where \(X_v\) is the
\(v\)-th views of the data. From the statistical point of view, we are
simply stacking observational models \(\mathcal{M}_v\) on each view,
with a conditional independence assumption with respect the common (and
unknown) partition:

\begin{equation}
\p(\Obs_1, \ldots, \Obs_V \mid \Clust) = \p(\Obs_1 \mid \mathcal{M}_1, \Clust) \times \ldots \times p(\Obs_V \mid \mathcal{M}_V, \Clust).
\end{equation}

Then, the \(\ICLex\) of the whole dataset is simply the sum of the
individual \(\ICLex\) of the V models, and the algorithms implemented in
\pkg{greed} still apply.

\hypertarget{the-combinedmodels-class}{%
\subsubsection{The CombinedModels
class}\label{the-combinedmodels-class}}

From an implementation point of view, all the combined models may have
their own hyperparameters \(\bBeta_v\), with the only constraint that
they must share the same partition \(\Clust\) and thus the same
(symmetric) Dirichlet prior with hyperparameter \(\alpha\) on the
cluster proportions \(\bPi\).

The \code{CombinedModels} S4 class implement this approach, and its
constructor takes two arguments

\begin{itemize}
\item \code{models} a named list of size $V$ containing S4 objects of class \code{DlvmPrior-class} instantiating each observational model and their hyperparameters $\bBeta_v$.
\item \code{alpha}  the cluster proportions the Dirichlet hyperparameter, same as before (default to 1)
\end{itemize}

Here, the \code{DlvmPrior-class} is used instead of the
\code{Dlvm-class}. It ensures that only the observational models
hyperparameters \(\bBeta_v\) can be specified, and not one \(\alpha_v\)
per model.

The data are provided with a named list, and the model is created with
the \code{?CombinedModels} constructor, which must also be provided with
a list of \code{models} priors, with the same \code{names} as in the
data list. The next cell gives the possible skeleton of a call to
\code{greed()} in this context, where may be replaced by any of the
implemented models described in \Cref{tab:obsmodS4}.

\begin{verbatim}
my_combined_mod <- CombinedModels(models = list("mod1"=<Mod1>Prior(beta1), 
                                                 ..., 
                                                 "modV"=<ModV>Prior(beta_V)),
                                  alpha=1)
X = list("mod1"=X_1, ..., "modV"=X_V)
sol = greed(X, model = my_combined_mod)
\end{verbatim}

Eventually, one can retrieve a fitted submodel on a specific view via
the \code{extractSubModel()} method and use all the classical methods of
individuals model on the result, as illustrated in the next code chunck.

\begin{verbatim}
submod_1 = extractSubModel(sol, "mod1") 
\end{verbatim}

We now illustrate the method on a concrete heterogeneous dataset mixing
continuous and categorical variables, with \code{GmmPrior()} and
\code{LcaPrior()} respectively stacked on each view. However, we
emphasize that a \code{CombinedModels} may be built with any combination
of models (including graph models), and that each model may appear
multiple times. For example, it can be used to create models for
multilayered networks and/or networks with nodes attributes
\footnote{For example of such complex models we refers the reader to the packages vignettes and in particular the \textbf{Graph Clustering} and \textbf{Combined Models} vignettes available online at \href{https://comeetie.github.io/greed/}{https://comeetie.github.io/greed/}.}.

\hypertarget{illustration-on-the-fifa-data-set}{%
\subsubsection{Illustration on the Fifa
data-set}\label{illustration-on-the-fifa-data-set}}

The \textbf{Fifa} dataset is attached to \pkg{greed} and contains
several features of soccer players in the Fifa 2020 videogame (see
\code{?Fifa}). After loading the dataset, we begin by a small
pre-processing only keeping players being worth more than 1 000 000
euros, while removing columns corresponding to the players' names,
nationalities, values, a boolean indicating if the player is a goal
keeper, and the player \(x\) and \(y\) coordinate on a 2-D view of the
field. These coordinates will be useful later when analysing the
clustering result.

\begin{CodeChunk}
\begin{CodeInput}
R> data("Fifa")
R> X <- Fifa[,-c(1,2)] %>% 
+   filter(value_eur>1000000) %>% 
+   select(-value_eur,-GK,-pos_x,-pos_y)
\end{CodeInput}
\end{CodeChunk}

This leaves us with a dataset of \(n=2266\) rows and \(p=24\) columns.
The features correspond to players' characteristics (such as age, height
or weight), skills statistics (pace, shooting, \ldots) and a set of
binary factors that indicates the possible field positions of each
player (\textit{e.g.} attack right or middle centre) as well as their
preferred foot.

The next code block demonstrates the \code{CombinedModels} approach and
its instantiation for this dataset. The data are split into categorical
and continuous \emph{views} with \code{LcaPrior()} and \code{GmmPrior()}
models respectively,

\begin{CodeChunk}
\begin{CodeInput}
R> data <- list(categorical = X %>% select_if(is.factor),
+             cont = X %>% select_if(function(v){!is.factor(v)}))
R> cbmods <- CombinedModels(models=list(categorical=LcaPrior(),cont=GmmPrior()))
R> sol = greed(data,model = cbmods)
\end{CodeInput}
\end{CodeChunk}

\begin{quote}
\textbf{Note:} Be careful, names in the models list and in the data list
must match.
\end{quote}

The resulting model being fitted, the \code{extractSubModel()} method
allows retrieving each sub-model, via its name in the \code{models}
list, and thus to use the plotting capabilities of \pkg{greed} on all
views. For example, \Cref{fig:genplotcat} investigates the categorical
part of the model and plot the marginal distribution of the factor
variables in each cluster thanks to the \code{marginals} plot type
implemented specifically for LCA. We see that the estimated clusters
agree quite strongly with the field positions features. The preferred
foot feature also have a clear impact on two clusters.

\begin{CodeChunk}
\begin{CodeInput}
R> plot(extractSubModel(sol,"categorical"),type="marginals")
\end{CodeInput}
\begin{figure}

{\centering \includegraphics[width=1\linewidth]{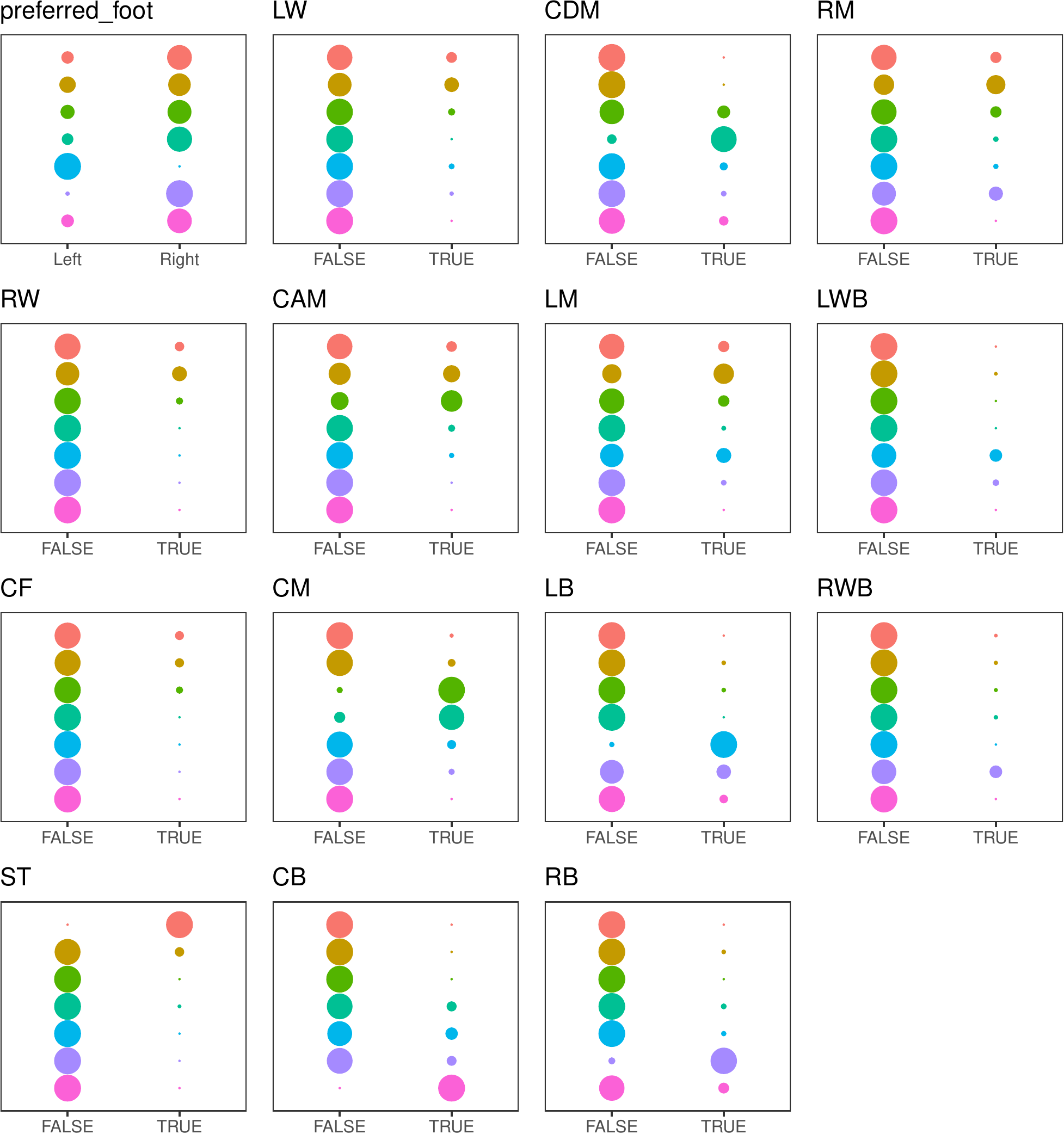} 

}

\caption[Marginal plots of the Lca part of the model over categorical features]{Marginal plots of the Lca part of the model over categorical features.}\label{fig:genplotcat}
\end{figure}
\end{CodeChunk}

For the continuous features, \Cref{fig:genplotviol} shows the
\code{violins} plot type specifically implemented for GMMs. Again, the
clusters seem well organized and the hierarchical ordering has produced
a meaningful ordering as can be seen with the shooting feature. We will
go back to this point later.

\begin{CodeChunk}
\begin{CodeInput}
R> plot(extractSubModel(sol,"cont"),type="violins")
\end{CodeInput}
\begin{figure}

{\centering \includegraphics[width=1\linewidth]{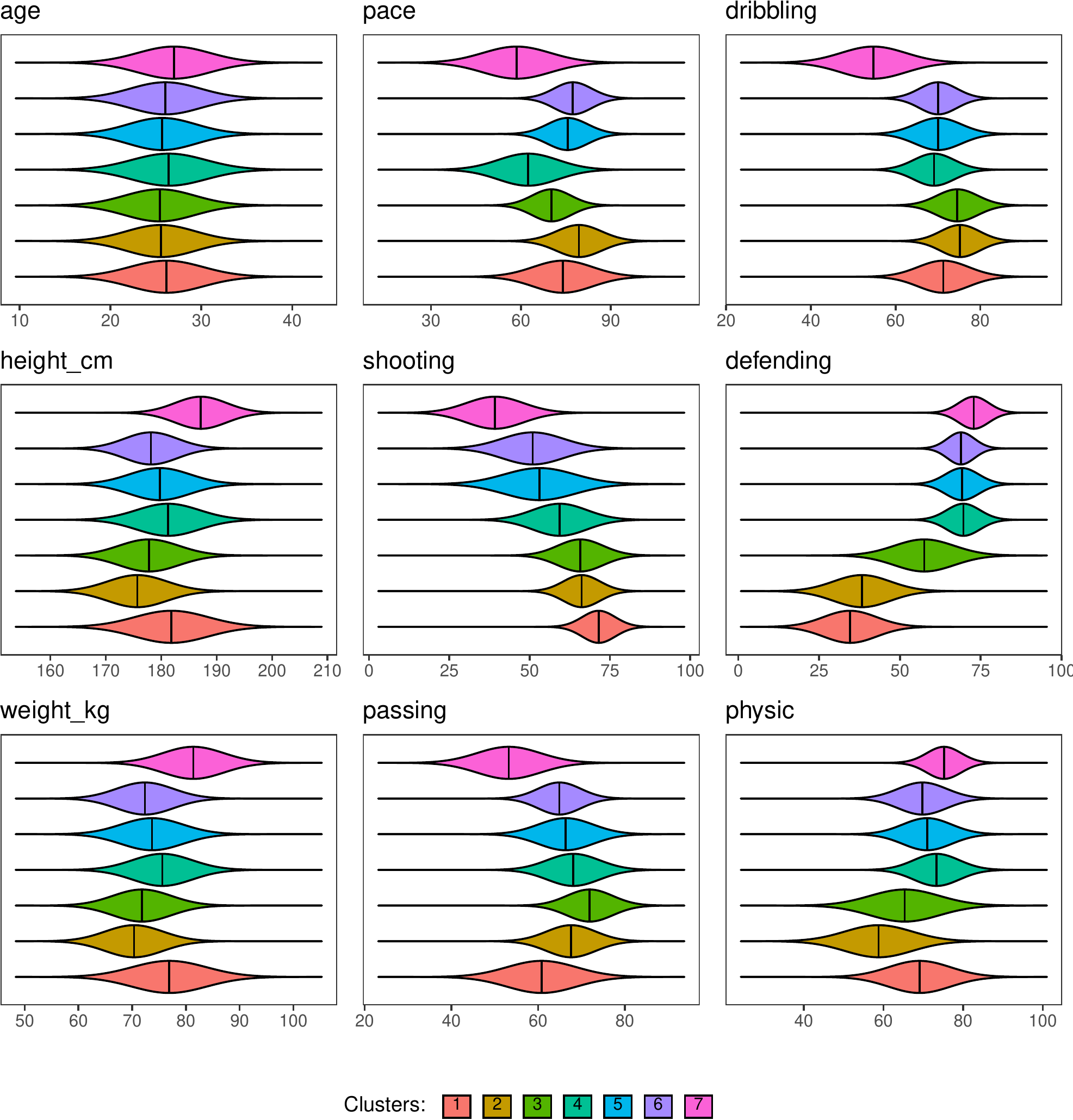} 

}

\caption[Violins plots of the Gmm part of the model over continuous features]{Violins plots of the Gmm part of the model over continuous features.}\label{fig:genplotviol}
\end{figure}
\end{CodeChunk}

\Cref{fig:genplotmar} shows the plot of \code{marginals} density plots
in each clusters.The bimodal aspect of the defending features clearly
appears in this visualization, while the mixture of the remaining
features is clearly poorly separated.

\begin{CodeChunk}
\begin{CodeInput}
R> plot(extractSubModel(sol,"cont"),type="marginals")
\end{CodeInput}
\begin{figure}

{\centering \includegraphics[width=1\linewidth]{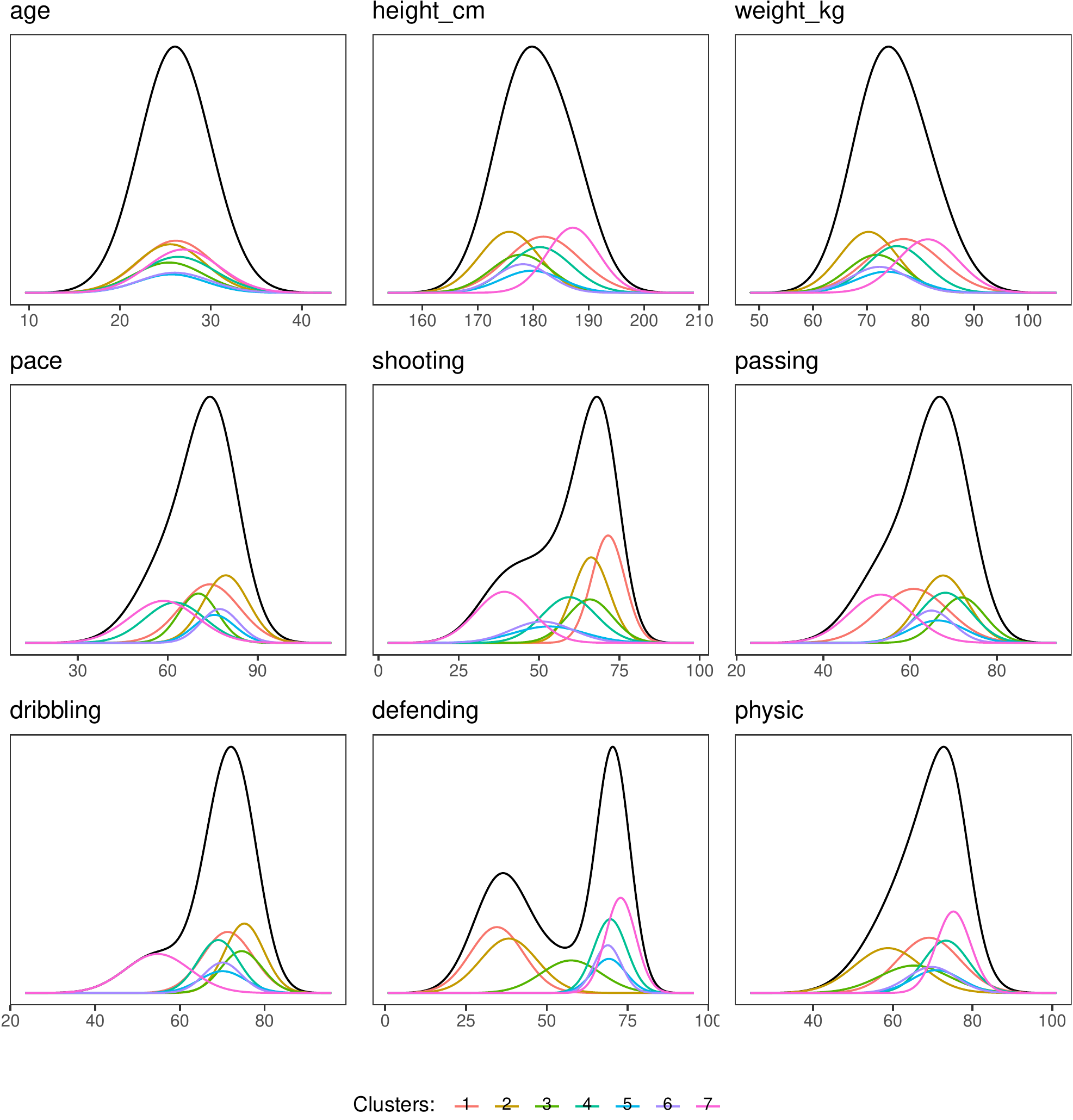} 

}

\caption[Marginal plots of the Gmm part of the model over continuous features]{Marginal plots of the Gmm part of the model over continuous features.}\label{fig:genplotmar}
\end{figure}
\end{CodeChunk}

Finally, in order to clearly show the alignment of the cluster with
field position, \Cref{fig:posplot} displays the average position on the
field for each cluster.

\begin{CodeChunk}
\begin{CodeInput}
R> clust_positions = Fifa[,-c(1,2)] %>% 
+   filter(value_eur>1000000) %>%
+   select(pos_x,pos_y) %>%
+   mutate(cluster=clustering(sol)) %>%
+   group_by(cluster) %>%
+   summarize(pos_x=mean(pos_x),pos_y=mean(pos_y))
R> 
R> ggplot(clust_positions)+
+   annotate_pitch()+
+   geom_text(aes(x=pos_x,y=pos_y,label=cluster),size=5,col="red")+
+   theme_pitch() 
\end{CodeInput}
\begin{figure}

{\centering \includegraphics[width=0.9\linewidth]{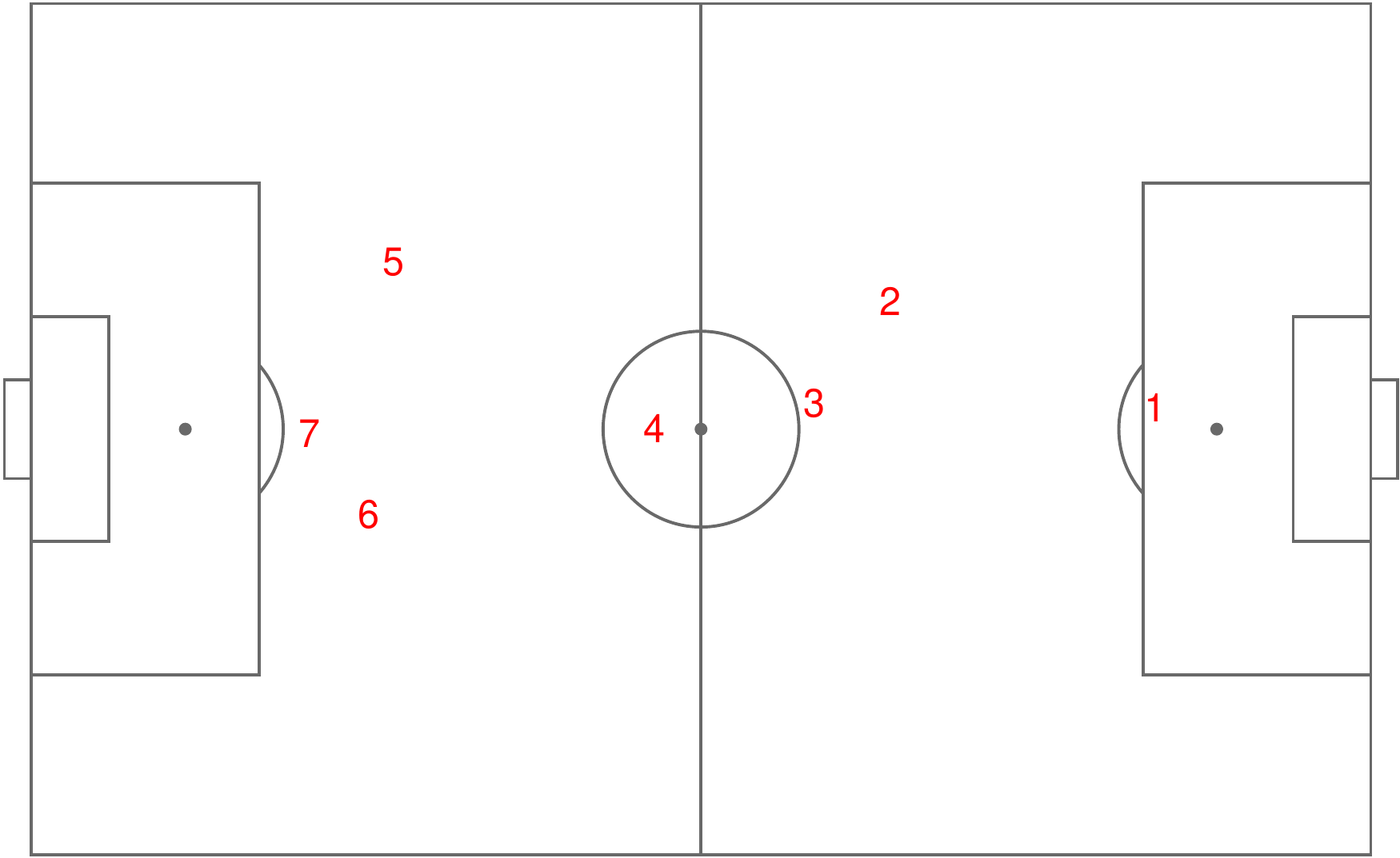} 

}

\caption[Field position map of the clusters]{Field position map of the clusters.}\label{fig:posplot}
\end{figure}
\end{CodeChunk}

The organization of the clusters clearly appears, cluster 7 correspond
to central defensive players, cluster 5 and 6 to lateral defensive
players (recall that the preferred foot of this player were quite
different between these two clusters with almost only right footed
players in cluster 6 and left footed players in cluster 5).

\hypertarget{conclusion}{%
\section{Conclusion}\label{conclusion}}

This paper introduced \pkg{greed}, a new \proglang{R} package for
model-based clustering with a flexible interface and a wide range of
applications from multivariate data to graphs and heterogeneous data. In
particular, the package implements two new algorithms. First, a hybrid
GA algorithm that performs both clustering and model-selection
altogether. Second, a hierarchical post-processing which allows cluster
ordering and extraction of coarser solutions, hence enabling multiscale
analysis of the clustering results. Combined, these two algorithms offer
a competitive approach in terms of computational time, ease of use and
results interpretability. The generic aspect of the approach is
reflected in the package implementation, with S4 class representation of
observational models and the possibility to implement new ones.
Moreover, the implementation of popular observational models allows to
handle a wide variety of datasets, of possibly heterogenous nature, as
illustrated on real datasets. We emphasize that the key advantage of the
hybrid-GA - enabling its genericity - is its random initialization,
making it independent from carefully chosen seeding procedures. Indeed,
the latter is both heavily model-dependent and may also have a strong
influence of the convergence of classical EM-like algorithms extensively
used in model-based clustering. To finish, while already versatile, we
lay several possible extensions to the package in order to widen its
scope. First, a desirable feature would be to allow for multiple
measurement data, without changing the generic interface. Second,
handling missing data is also a desirable feature in modern
applications. The DLVM framework allows for a generic approach in two
contexts: either with a missing at random hypothesis, or when the
missing pattern is supposed to only depends on \(\Clust\). Finally,
handling models for which there exists no explicit formula of the
\(\ICLex\) criterion is a natural direction. However, it would required
the adaptation of the two main algorithms to handle the approximation
steps and is therefore postpone to future works.

\bibliography{biblioJSS.bib}

\begin{thebibliography}{40}
\newcommand{\enquote}[1]{``#1''}
\providecommand{\natexlab}[1]{#1}
\providecommand{\url}[1]{\texttt{#1}}
\providecommand{\urlprefix}{URL }
\expandafter\ifx\csname urlstyle\endcsname\relax
  \providecommand{\doi}[1]{doi:\discretionary{}{}{}#1}\else
  \providecommand{\doi}{doi:\discretionary{}{}{}\begingroup
  \urlstyle{rm}\Url}\fi
\providecommand{\eprint}[2][]{\url{#2}}

\bibitem[{Akaike(1974)}]{akaike1974new}
Akaike H (1974).
\newblock \enquote{A new look at the statistical model identification.}
\newblock \emph{IEEE transactions on automatic control}, \textbf{19}(6),
  716--723.

\bibitem[{Bar-Joseph \emph{et~al.}(2001)Bar-Joseph, Gifford, and
  Jaakkola}]{Bar2001}
Bar-Joseph Z, Gifford DK, Jaakkola TS (2001).
\newblock \enquote{{Fast optimal leaf ordering for hierarchical clustering }.}
\newblock \emph{Bioinformatics}, \textbf{17}(1), S22--S29.
\newblock ISSN 1367-4803.
\newblock
  \eprint{http://oup.prod.sis.lan/bioinformatics/article-pdf/17/suppl\_1/S22/726790/17S022.pdf},
  \urlprefix\url{https://doi.org/10.1093/bioinformatics/17.suppl\_1.S22}.

\bibitem[{Bates and Maechler(2019)}]{Bates2019}
Bates D, Maechler M (2019).
\newblock \emph{Matrix: Sparse and Dense Matrix Classes and Methods}.
\newblock R package version 1.2-17,
  \urlprefix\url{https://CRAN.R-project.org/package=Matrix}.

\bibitem[{Benaglia \emph{et~al.}(2009)Benaglia, Chauveau, Hunter, and
  Young}]{mixtools}
Benaglia T, Chauveau D, Hunter DR, Young D (2009).
\newblock \enquote{{mixtools}: An {R} Package for Analyzing Finite Mixture
  Models.}
\newblock \emph{Journal of Statistical Software}, \textbf{32}(6), 1--29.
\newblock \urlprefix\url{http://www.jstatsoft.org/v32/i06/}.

\bibitem[{Bengtsson(2019)}]{Bengtsson2019}
Bengtsson H (2019).
\newblock \emph{future: Unified Parallel and Distributed Processing in R for
  Everyone}.
\newblock R package version 1.13.0,
  \urlprefix\url{https://CRAN.R-project.org/package=future}.

\bibitem[{Bertoletti \emph{et~al.}(2015)Bertoletti, Friel, and
  Rastelli}]{Bertoletti2015}
Bertoletti M, Friel N, Rastelli R (2015).
\newblock \enquote{Choosing the number of clusters in a finite mixture model
  using an exact integrated completed likelihood criterion.}
\newblock \emph{METRON}, \textbf{73}(2), 177--199.
\newblock ISSN 2281-695X.
\newblock \doi{10.1007/s40300-015-0064-5}.
\newblock \urlprefix\url{https://doi.org/10.1007/s40300-015-0064-5}.

\bibitem[{Biernacki \emph{et~al.}(2000)Biernacki, Celeux, and
  Govaert}]{Biernacki2000}
Biernacki C, Celeux G, Govaert G (2000).
\newblock \enquote{Assessing a mixture model for clustering with the integrated
  completed likelihood.}
\newblock \emph{IEEE Transaction on Pattern Analysis and Machine Intelligence},
  \textbf{7}, 719--725.

\bibitem[{Blondel \emph{et~al.}(2008)Blondel, Guillaume, Lambiotte, and
  Lefebvre}]{blondel2008fast}
Blondel VD, Guillaume JL, Lambiotte R, Lefebvre E (2008).
\newblock \enquote{Fast unfolding of communities in large networks.}
\newblock \emph{Journal of statistical mechanics: theory and experiment},
  \textbf{2008}(10), P10008.

\bibitem[{Bouveyron \emph{et~al.}(2019)Bouveyron, Celeux, Murphy, and
  Raftery}]{bouveyron2019model}
Bouveyron C, Celeux G, Murphy TB, Raftery AE (2019).
\newblock \emph{Model-Based Clustering and Classification for Data Science:
  With Applications in R}, volume~50.
\newblock Cambridge University Press.

\bibitem[{Chiquet \emph{et~al.}(2021)Chiquet, Donnet, and Barbillon}]{sbm}
Chiquet J, Donnet S, Barbillon P (2021).
\newblock \emph{sbm: Stochastic Blockmodels}.
\newblock R package version 0.4.3,
  \urlprefix\url{https://CRAN.R-project.org/package=sbm}.

\bibitem[{C{\^o}me \emph{et~al.}(2021)C{\^o}me, Jouvin, Latouche, and
  Bouveyron}]{come2021hierarchical}
C{\^o}me E, Jouvin N, Latouche P, Bouveyron C (2021).
\newblock \enquote{Hierarchical clustering with discrete latent variable models
  and the integrated classification likelihood.}
\newblock \emph{Advances in Data Analysis and Classification}, pp. 1--30.

\bibitem[{C\^{o}me and Latouche(2015)}]{come2015}
C\^{o}me E, Latouche P (2015).
\newblock \enquote{Model selection and clustering in stochastic block models
  based on the exact integrated complete data likelihood.}
\newblock \emph{Statistical Modelling}, \textbf{15}(6), 564--589.

\bibitem[{Csardi and Nepusz(2006)}]{igraph}
Csardi G, Nepusz T (2006).
\newblock \enquote{The igraph software package for complex network research.}
\newblock \emph{InterJournal}, \textbf{Complex Systems}, 1695.
\newblock \urlprefix\url{https://igraph.org}.

\bibitem[{Eddelbuettel and Balamuta(2017)}]{Eddelbuettel2017}
Eddelbuettel D, Balamuta JJ (2017).
\newblock \enquote{{Extending extit{R} with extit{C++}: A Brief Introduction to
  extit{Rcpp}}.}
\newblock \emph{PeerJ Preprints}, \textbf{5}, e3188v1.
\newblock ISSN 2167-9843.
\newblock \doi{10.7287/peerj.preprints.3188v1}.
\newblock \urlprefix\url{https://doi.org/10.7287/peerj.preprints.3188v1}.

\bibitem[{Eddelbuettel and Sanderson(2014)}]{Eddelbuettel2014}
Eddelbuettel D, Sanderson C (2014).
\newblock \enquote{RcppArmadillo: Accelerating R with high-performance C++
  linear algebra.}
\newblock \emph{Computational Statistics and Data Analysis}, \textbf{71},
  1054--1063.
\newblock \urlprefix\url{http://dx.doi.org/10.1016/j.csda.2013.02.005}.

\bibitem[{Eiben and Smith(2004)}]{Eiben2003}
Eiben AE, Smith JE (2004).
\newblock \emph{Introduction to Evolutionary Computing, 2$^{nd}$ Edition}.
\newblock Springer-Verlag.

\bibitem[{Everitt \emph{et~al.}(2011)Everitt, Landau, and
  Leese}]{everitt2011cluster}
Everitt BS, Landau S, Leese M (2011).
\newblock \emph{Cluster Analysis, Fifth Edition (Wiley Series in Probability
  and Statistics)}.
\newblock Wiley Series in Probability and Statistics, 5th edition. Wiley.
\newblock ISBN
  0470749911,9780470749913,9780470977804,9780470977811,9780470978443.

\bibitem[{Fruhwirth-Schnatter \emph{et~al.}(2019)Fruhwirth-Schnatter, Celeux,
  and Robert}]{fruhwirth2019handbook}
Fruhwirth-Schnatter S, Celeux G, Robert CP (2019).
\newblock \emph{Handbook of mixture analysis}.
\newblock Chapman and Hall/CRC.

\bibitem[{Gr\"un and Leisch(2008)}]{flexmix}
Gr\"un B, Leisch F (2008).
\newblock \enquote{{FlexMix} Version 2: Finite Mixtures with Concomitant
  Variables and Varying and Constant Parameters.}
\newblock \emph{Journal of Statistical Software}, \textbf{28}(4), 1--35.
\newblock \doi{10.18637/jss.v028.i04}.
\newblock \urlprefix\url{https://www.jstatsoft.org/v28/i04/}.

\bibitem[{{Hruschka} \emph{et~al.}(2009){Hruschka}, {Campello}, {Freitas}, and
  {Ponce Leon F. de Carvalho}}]{Hruschka2009}
{Hruschka} ER, {Campello} RJGB, {Freitas} AA, {Ponce Leon F de Carvalho} AC
  (2009).
\newblock \enquote{A Survey of Evolutionary Algorithms for Clustering.}
\newblock \emph{IEEE Transactions on Systems, Man, and Cybernetics, Part C
  (Applications and Reviews)}, \textbf{39}(2), 133--155.
\newblock \doi{10.1109/TSMCC.2008.2007252}.

\bibitem[{Karatzoglou \emph{et~al.}(2004)Karatzoglou, Smola, Hornik, and
  Zeileis}]{kernlab}
Karatzoglou A, Smola A, Hornik K, Zeileis A (2004).
\newblock \enquote{kernlab -- An {S4} Package for Kernel Methods in {R}.}
\newblock \emph{Journal of Statistical Software}, \textbf{11}(9), 1--20.
\newblock \urlprefix\url{http://www.jstatsoft.org/v11/i09/}.

\bibitem[{Langrognet \emph{et~al.}(2020)Langrognet, Lebret, Poli, Iovleff,
  Auder, and Iovleff}]{Rmixmod}
Langrognet F, Lebret R, Poli C, Iovleff S, Auder B, Iovleff S (2020).
\newblock \emph{Rmixmod: Classification with Mixture Modelling}.
\newblock R package version 2.1.5,
  \urlprefix\url{https://CRAN.R-project.org/package=Rmixmod}.

\bibitem[{Leger \emph{et~al.}(2020)Leger, Barbillon, and Chiquet}]{blockmodels}
Leger JB, Barbillon P, Chiquet J (2020).
\newblock \emph{blockmodels: Latent and Stochastic Block Model Estimation by a
  'V-EM' Algorithm}.
\newblock R package version 1.1.4,
  \urlprefix\url{https://CRAN.R-project.org/package=blockmodels}.

\bibitem[{Linzer and Lewis(2011)}]{poLCA}
Linzer DA, Lewis JB (2011).
\newblock \enquote{{poLCA}: An {R} Package for Polytomous Variable Latent Class
  Analysis.}
\newblock \emph{Journal of Statistical Software}, \textbf{42}(10), 1--29.
\newblock \urlprefix\url{http://www.jstatsoft.org/v42/i10/}.

\bibitem[{MacQueen(1967)}]{macqueen1967some}
MacQueen J (1967).
\newblock \enquote{Some methods for classification and analysis of multivariate
  observations.}
\newblock In \emph{Proceedings of the fifth Berkeley symposium on mathematical
  statistics and probability}, volume~1, pp. 281--297. Oakland, CA, USA.

\bibitem[{Maechler \emph{et~al.}(2021)Maechler, Rousseeuw, Struyf, Hubert, and
  Hornik}]{clusterpackage}
Maechler M, Rousseeuw P, Struyf A, Hubert M, Hornik K (2021).
\newblock \emph{cluster: Cluster Analysis Basics and Extensions}.
\newblock R package version 2.1.2 --- For new features, see the 'Changelog'
  file (in the package source),
  \urlprefix\url{https://CRAN.R-project.org/package=cluster}.

\bibitem[{Mohammadi(2021)}]{bmixture}
Mohammadi R (2021).
\newblock \emph{bmixture: Bayesian Estimation for Finite Mixture of
  Distributions}.
\newblock R package version 1.7.

\bibitem[{Mouselimis(2021)}]{ClusterR}
Mouselimis L (2021).
\newblock \emph{{ClusterR}: Gaussian Mixture Models, K-Means,
  Mini-Batch-Kmeans, K-Medoids and Affinity Propagation Clustering}.
\newblock R package version 1.2.5,
  \urlprefix\url{https://CRAN.R-project.org/package=ClusterR}.

\bibitem[{Murphy \emph{et~al.}(2021)Murphy, Viroli, and Gormley}]{IMIFA}
Murphy K, Viroli C, Gormley IC (2021).
\newblock \emph{\texttt{\textup{IMIFA}}: {I}nfinite Mixtures of Infinite Factor
  Analysers and Related Models}.
\newblock \textsf{R} package version 2.1.6,
  \urlprefix\url{https://cran.r-project.org/package=IMIFA}.

\bibitem[{Ng \emph{et~al.}(2002)Ng, Jordan, and Weiss}]{ng2002spectral}
Ng AY, Jordan MI, Weiss Y (2002).
\newblock \enquote{On spectral clustering: Analysis and an algorithm.}
\newblock In \emph{Advances in neural information processing systems}, pp.
  849--856.

\bibitem[{Pedregosa \emph{et~al.}(2011)Pedregosa, Varoquaux, Gramfort, Michel,
  Thirion, Grisel, Blondel, Prettenhofer, Weiss, Dubourg, Vanderplas, Passos,
  Cournapeau, Brucher, Perrot, and Duchesnay}]{scikit-learn}
Pedregosa F, Varoquaux G, Gramfort A, Michel V, Thirion B, Grisel O, Blondel M,
  Prettenhofer P, Weiss R, Dubourg V, Vanderplas J, Passos A, Cournapeau D,
  Brucher M, Perrot M, Duchesnay E (2011).
\newblock \enquote{Scikit-learn: Machine Learning in {P}ython.}
\newblock \emph{Journal of Machine Learning Research}, \textbf{12}, 2825--2830.

\bibitem[{Peixoto(2014)}]{graphtool}
Peixoto TP (2014).
\newblock \enquote{The graph-tool python library.}
\newblock \emph{figshare}.
\newblock \doi{10.6084/m9.figshare.1164194}.
\newblock \urlprefix\url{http://figshare.com/articles/graph_tool/1164194}.

\bibitem[{Plummer \emph{et~al.}(2019)Plummer, Stukalov, and Denwood}]{rjags}
Plummer M, Stukalov A, Denwood M (2019).
\newblock \enquote{{rjags}: Bayesian Graphical Models using MCMC.}

\bibitem[{{R Core Team}(2021)}]{Rcore}
{R Core Team} (2021).
\newblock \emph{R: A Language and Environment for Statistical Computing}.
\newblock R Foundation for Statistical Computing, Vienna, Austria.
\newblock \urlprefix\url{https://www.R-project.org/}.

\bibitem[{Robert and Casella(2013)}]{robert2013monte}
Robert C, Casella G (2013).
\newblock \emph{Monte Carlo statistical methods}.
\newblock Springer Science \& Business Media.

\bibitem[{Schwarz(1978)}]{schwarz1978estimating}
Schwarz G (1978).
\newblock \enquote{Estimating the dimension of a model.}
\newblock \emph{The annals of statistics}, \textbf{6}(2), 461--464.

\bibitem[{Scrucca \emph{et~al.}(2016)Scrucca, Fop, Murphy, and
  Raftery}]{Mclust}
Scrucca L, Fop M, Murphy TB, Raftery AE (2016).
\newblock \enquote{{mclust} 5: clustering, classification and density
  estimation using {G}aussian finite mixture models.}
\newblock \emph{The {R} Journal}, \textbf{8}(1), 289--317.
\newblock \urlprefix\url{https://doi.org/10.32614/RJ-2016-021}.

\bibitem[{{Stan Development Team}(2020)}]{RStan}
{Stan Development Team} (2020).
\newblock \enquote{{RStan}: the {R} interface to {Stan}.}
\newblock R package version 2.21.2, \urlprefix\url{http://mc-stan.org/}.

\bibitem[{Szepannek(2018)}]{clustMixType}
Szepannek G (2018).
\newblock \enquote{clustMixType: User-Friendly Clustering of Mixed-Type Data in
  R.}
\newblock \emph{The R Journal}, pp. 200--208.
\newblock \doi{10.32614/RJ-2018-048}.
\newblock \urlprefix\url{https://doi.org/10.32614/RJ-2018-048}.

\bibitem[{Taddy(2016)}]{Bmix}
Taddy M (2016).
\newblock \enquote{Bmix: Bayesian Sampling for Stick-Breaking Mixtures.}

\end{thebibliography}

\end{document}